\documentclass[12pt]{article}
\usepackage[utf8]{inputenc}

\usepackage[top=1in, bottom=1in, left=1in, right=1in]{geometry}
\usepackage{graphicx} 
\usepackage{subfig}
\usepackage{amsthm}
\usepackage[comma,numbers]{natbib}
\usepackage{ulem}
\usepackage{color}
\usepackage{lineno}  
\usepackage{hyperref}
\usepackage{amsmath}
\usepackage{amssymb}
\usepackage{bm}
\usepackage{indentfirst} 
\usepackage{mathrsfs}
\usepackage{algorithm} 
\usepackage{algorithmic}
\usepackage{threeparttable}
\usepackage{booktabs}
\numberwithin{equation}{section}

\newcommand\keywords[1]{\textbf{Keywords}: #1}

\title{The pseudo-analytical density solution to parameterized Fokker-Planck equations via deep learning}

\author{\small{Xiaolong Wang$^{1,2,3}$, Jing Feng$^4$, Gege Wang$^2$, Tong Li$^1$, Yong Xu$^{2,3}$\footnote{Corresponding author. E-mail addresses:  {\it hsux3@nwpu.edu.cn} (Y. Xu)}} \\
\small{$^1$ School of Mathematics and Statistics,}\\ \small{Shaanxi Normal University, Xi'an, 710119, China}\\
\small{$^2$ School of Mathematics and Statistics,}\\ \small{Northwestern Polytechnical University, Xi'an, 710129, China}\\
\small{$^3$ MOE Key Laboratory for Complexity Science in Aerospace,}\\ \small{Northwestern Polytechnical University, Xi'an, 710072, China}\\
\small{$^4$ School of Science, Xi'an University of Posts and Telecommunications,}\\ \small{Xi'an, 710121, China}
}

\date{}

\begin{document}

\maketitle
 

\begin{abstract} 

Efficiently solving the Fokker-Planck equation (FPE) is crucial for understanding the probabilistic evolution of stochastic particles in dynamical systems, however, analytical solutions or density functions are only attainable in specific cases. To speed up the solving process of parameterized FPEs with several system parameters, we introduce a deep learning-based method to obtain the pseudo-analytical density (PAD). Unlike previous numerical methodologies that necessitate solving the FPE separately for each set of system parameters, the PAD simultaneously addresses all the FPEs within a predefined continuous range of system parameters during a single training phase. The approach utilizes a Gaussian mixture distribution (GMD) to represent the stationary probability density, the solution to the FPE. By leveraging a deep residual network, each system parameter configuration is mapped to the parameters of the GMD, ensuring that the weights, means, and variances of the Gaussian components adaptively align with the corresponding true density functions. A grid-free algorithm is further developed to effectively train the residual network, resulting in a feasible PAD obeying necessary normalization and boundary conditions. Extensive numerical studies validate the accuracy and efficiency of our method, promising significant acceleration in the response analysis of multi-parameter, multi-dimensional stochastic nonlinear systems.
\bigskip

\noindent\keywords{nonlinear stochastic system, Fokker-Planck equation, analytical solution, deep learning, Gaussian mixture distribution.}
\end{abstract}

\section{Introduction}
The Fokker-Planck equation (FPE), named after Adriaan Fokker and Max Planck, is a fundamental partial differential equation (PDE) for studying stochastic processes. It describes the time evolution of the probability density function (PDF) of the particles governed by a stochastic differential equation and establishes a powerful research perspective, i.e., studying stochastic behaviors by solving PDEs. The transient and stationary solutions of the FPE thus provide solid probabilistic information about stochastic systems, prompting extensive research into solving specific FPEs. As the analytical density expression only exist on some low-dimensional or linear systems, numerically solving FPEs has become an active research area. Existing approaches include finite difference method~\cite{Neena2022Some, Kumar2006Sadhana}, finite element method~\cite{Pichler2013Numerical, Naprstek2014Finite}, path integral~\cite{Drozdov1998Accurate,Xu2019Path}, generalized cell mapping~\cite{Yue2019Probabilistic}, Monte Carlo simulation~\cite{Muscolino2003Muscolino}, and machine learning methods~\cite{Xu2020Solving, Wang2022Random}. However, these approaches are currently restricted to computing one system at a time. Whenever the system parameters undergo changes, recalculation is required. In numerous applications, the stochastic behavior of the system hinges on one or several control parameters. If the solving processes of the FPEs across all relevant system parameters could be parallelized in a single solving session, the response analysis of parameter-dependent stochastic systems will be greatly accelerated.

To this end, we introduce the pseudo-analytical density (PAD), a numerical solution that jointly solves parameterized FPEs from a single solving process. A PAD is an atlas of FPE solutions with tunable system parameters, from which one can immediately obtain the stationary PDF (SPDF) with any system parameters in the multi-dimensional parameter domain. It is evident that the map from the parameters of stochastic nonlinear systems to the SPDFs is very intricate. For instance, changing the parameters of the deterministic system and noise fluctuation may trigger stochastic P-bifurcation and D-bifurcation~\cite{Xu2011Hopf,Zhang2019Stochastic}. To build the PAD, we tune to machine learning methodologies, which have shown great success in modelling complex mappings and functions~\cite{Redmon2016yolo,Wang2024Noise,Yu2024Multi}.

In contrast to traditional methods, machine learning methods such as the physics-informed neural network (PINN)~\cite{Raissi2019pinn,OLeary2022Stochastic} represent the solution of differential equations by a neural network. As the memory-intensive grid or mesh of the state domain is avoid, neural networks can efficiently learn multi-dimensional systems. Consequently, the solving task is replaced by an optimization process that minimizes a suitable loss function, which jointly considers the differential equation, initial condition, and boundary condition. Equation solving thus consists of two main procedures: (1) building a suitable network architecture capable of approximating the solution and (2) designing an appropriate learning algorithm to optimize the neural network. Latest deep learning-based studies have explored the concurrent learning of the vector fields~\cite{Wang2024Deep} and solutions~\cite{Cho2024Parameterized,Liu2024Parameterized} of parameterized differential equations by a single deep neural network with a large number of layers. However, the constructions in~\cite{Cho2024Parameterized,Liu2024Parameterized} are not suitable to represent the solutions of the FPEs, which are PDFs. An efficient learning model should not only be capable of simultaneously representing the solutions with different system parameters, but also imposing the constraints of PDFs, especially the normalization condition.

At present, there are two approaches to solve the FPE and deal with the normalization condition. Xu et al.~\cite{Xu2020Solving,Zhang2020Statistical} have used a fully connected neural network (FCNN) to model the solution of the FPE and included the normalization condition in the loss function. As the integral involved in the condition is approximated by a summation on the whole state domain, the calculation requires a great amount of integral points on a regular grid, whose number grows exponentially with the system dimension. To decrease the number of points, deep KD trees have bee utilized to construct an irregular but more efficient partition to cover the state domain~\cite{Zhang2022Solving}, and low-rank separation representation has been applied to simplify high-dimensional distributions by combining low-dimensional ones~\cite{Zhang2023Deep}. However, these amendments do not fundamentally overcome the problem that the architecture of the FCNN is too free to represent a PDF, such that the optimization process has to inefficiently enforce the normalization condition.

Alternatively, fitting density functions by mixture distributions, especially Gaussian-shaped functions, is a popular methodology~\cite{Ososkov2000Gaussian,Eckart2018HGMR}. However, it often involves a complex optimization process~\cite{Vishwajeet2018Adaptive}. The radial basis function neuron network (RBFNN)~\cite{Wang2022Random,Ye2022RBFNN,Wang2023Radial,Qian2023Solving,Wang2023On,Qian2024candidate} uses a convex combination of Gaussian distributions to approximate the solution of the FPE. The calculation of the normalization condition is reduced from the dense points in the state domain to the weights of the Gaussian components, which can be solved efficiently as the numerical integration is avoided. Nevertheless, the potential modelling capability of Gaussian mixture distribution (GMD) has not be fully exploited. The RBFNN only optimizes the weights of the Gaussian components, while their variances are predefined constants, and the mean values are fixed on a regular grid. Consequently, the RBFNN uses an enormous number of Gaussian components, limiting its application on multi-dimensional systems. Intuitively, the SPDF of a bistable system with two peaks~\cite{Xu2015Phase}, regardless of the system dimension, can be roughly but reasonably approximated by a combination of two Gaussian distributions, as long as the combination weights, the mean vectors and the covariance matrices can be adjusted to fit the two peaks~\cite{Gournelos2020Fitting}. Therefore, if the full parameters of the GMD can be jointly optimized, the number of Gaussian components for learning distributions could be greatly reduced.

In this work, we propose a deep learning-based approach, that a single training session concurrently solves the parameterized FPEs. The resulting PAD can generate the solutions of the FPEs with any parameter choices in the predefined parameter domain with high speed and accuracy. The main idea is to represent the solution of the FPE by a mixture distribution, i.e., a GMD~\cite{Eckart2018HGMR,Wang2022Random}, that is capable of modelling a broad range of distributions, and learn a deep learning-based parameter transform that maps every choice of system parameters to the parameters of the GMD. The combination weights, mean vectors, and covariance matrices of the Gaussian components evolve in accordance with the respective solutions of the FPEs. Furthermore, to make the optimization feasible and simple, the normalization condition is encoded in the network architecture and removed from the loss function. These configurations make our method implemented easily and suitable for solving multi-dimensional multi-parameter FPEs simultaneously.

The rest of the paper is organized as follows. Section~\ref{sec:2} introduces the background of the FPE solving problem. Section~\ref{sec:3} presents our deep learning method. Section~\ref{sec:num_ex} details the numerical experiments on several paradigmatic stochastic systems and Sec.~\ref{sec:discussion} discusses the training detail and the effectiveness of the proposed method. Section~\ref{sec:6} concludes the work.

\section{The Fokker-Planck equation}
\label{sec:2}
We consider the autonomous stochastic system
\begin{equation}\label{eq:sde}
\dot{\mathbf{x}}(t)=\mathbf{A}(\mathbf{x};\mathbf{\Theta}) + \mathbf{B}(\mathbf{\Theta})\mathbf{\Xi}(t),
\end{equation}
where $\mathbf{x}(t)=[x_{1}(t),\cdots,x_{D_\text{STA}}(t)]^\top\in\mathbb{R}^{D_\text{STA}}$ is the $D_\text{STA}$-dimensional ($D_\text{STA}$-D) state vector at time $t$, and the dot over variables represents derivatives with represent to $t$. The drift term $\mathbf{A}(\mathbf{x};\mathbf{\Theta})=[a_{1},\cdots,a_{D_\text{STA}}]^\top$ and the constant diffusion term $\mathbf{B}=[b_{ij}]\in\mathbb{R}^{D_\text{STA}\times D_\text{STA}}$ are controlled by $D_{\text{PAR}}$ system parameters $\mathbf{\Theta}=[\theta_1,\cdots,\theta_{D_{\text{PAR}}}]^\top$. $\mathbf{\Xi}(t)=[\xi_{1}(t),\cdots,\xi_{D_\text{STA}}(t)]^\top$ is a vector of standard white Gaussian noises (SWGNs), i.e., the components are mutually independent SWGNs with the autocorrelation function being the delta function. As Eq.~(\ref{eq:sde}) is a Markov process, the solution $p(\mathbf{x},t;\mathbf{\Theta})$ at time $t$ (the transient PDF) satisfies the FPE~\cite{Risken1996FP}
\begin{equation}\label{eq:fpe}
\frac{\partial p(\mathbf{x}, t;\mathbf{\Theta})}{\partial t}=\mathcal{L}_{\text{FP}} p(\mathbf{x}, t;\mathbf{\Theta}),
\end{equation}
where the FP operator is defined by
\begin{equation}
\mathcal{L}_{\text{FP}}p(\mathbf{x},t;\mathbf{\Theta}) = -\sum_{i=1}^{D_{\text{STA}}}\frac{\partial }{\partial x_{i}}\left[a_i(\mathbf{x};\mathbf{\Theta})\cdot p(\mathbf{x},t;\mathbf{\Theta})\right]+\frac{1}{2}\sum_{i=1}^{D_{\text{STA}}}\sum_{j=1}^{D_{\text{STA}}}\frac{\partial^2}{\partial x_i\partial x_j}\left[d_{ij}\cdot p(\mathbf{x},t;\mathbf{\Theta})\right],
\end{equation}
where $d_{ij}$ is the $ij$-entry of the diffusion matrix $\mathbf{D}=\mathbf{B}\mathbf{B}^\text{T}$, and both the functions $\{a_i\}_{i=1}^{D_\text{STA}}$ and the constants $\{d_{ij}\}_{i,j=1}^{D_\text{STA}}$ depend on the system parameters $\mathbf{\Theta}$. In particular, the SPDF $p(\mathbf{x};\mathbf{\Theta})$ satisfies the FPE
\begin{equation}\label{eq:condition_fp}
\mathcal{L}_{\text{FP}}p(\mathbf{x};\mathbf{\Theta})=0,
\end{equation}
which is the learning target of the presented work.

Apart from the FPE condition Eq.~(\ref{eq:condition_fp}), as $p(\mathbf{x};\mathbf{\Theta})$ is a PDF, it should also satisfy the nonnegativity condition of probability
\begin{equation}\label{eq:condition_nonega}
p(\mathbf{x};\mathbf{\Theta})\geq 0,
\end{equation}
for any state $\mathbf{x}\in\mathbb{R}^{D_\text{STA}}$, and the normalization condition of probability
\begin{equation}\label{eq:condition_norm}
\int_{\mathbb{R}^{D_\text{STA}}} p(\mathbf{x};\mathbf{\Theta})d\mathbf{x}=1.
\end{equation}
We only consider the natural boundary condition
\begin{equation}\label{eq:condition_boundary}
p(\mathbf{x};\mathbf{\Theta})=0 \quad\text{for}\quad\mathbf{x} \in \partial \mathcal{S},
\end{equation}
where $\partial \mathcal{S}$ is the boundary of the user-defined compact state domain $\mathcal{S} \subset \mathbb{R}^{D_\text{STA}}$, denoted by the $D_{\text{STA}}$-D box
\begin{equation}\label{eq:state_domain}
\mathcal{S}=\{\mathbf{x}|x_i\in[x_i^\text{min}, x_i^\text{max}], i=1,\cdots,D_\text{STA}\}.
\end{equation}

The goal of this work is to build a PAD $q(\mathbf{x};\mathbf{\Theta})$ to simultaneously approximate the SPDFs $p(\mathbf{x};\mathbf{\Theta})$ at all states $\mathbf{x}\in \mathcal{S}$ and all system parameters $\mathbf{\Theta}\in\mathcal{P}$, where $\mathcal{P}$ is the user-defined feasible parameter domain, denoted by the $D_\text{PAR}$-D box
\begin{equation}\label{eq:param_domain}
\mathcal{P}=\{\mathbf{\Theta}|\theta_i\in[\theta_i^\text{min},\theta_i^\text{max}],i=1,\cdots,D_\text{PAR}\}.
\end{equation}

\section{The deep mixture density network}
\label{sec:3}

To concurrently approximate the parameter-dependent PDFs $p(\mathbf{x};\mathbf{\Theta})$ for all $\mathbf{x}\in\mathcal{S}$ and $\mathbf{\Theta}\in\mathcal{P}$ by a PAD, two key considerations are raised: (i) how to construct a learning architecture that can simultaneously model a broad range of PDFs with necessary constraints, and (ii) how to successfully optimize such learning models. We prefer not to use a vanilla neural network, such as the FCNN, to model the PAD, because constraints such as the normalization condition must be inefficiently incorporated into the optimization process, forcing the neural network to output PDFs. Instead, it is recommended to encode the necessary constraints into the learning architecture such that it always outputs feasible density functions, which would significantly simplify and speed up the following optimization process.

We model the PAD $q(\mathbf{x};\mathbf{\Theta})$ by a mixture density network (MDN)~\cite{bishop1994mixture,He2020Deep,Chen2022Physics,Yu2024Multi} $q(\mathbf{x};\mathbf{\Theta})=q_{\text{GMD}}(\mathbf{x};\mathbf{\Phi})$, a GMD whose parameters $\mathbf{\Phi}$ are obtained by a neural network-based parameter transform $\mathbf{\Phi}=f(\mathbf{\Theta})$. The GMD can model multi-dimensional multi-modal distributions and automatically satisfies the nonnegativity condition of probability. It is then promising to construct a parameter transform $\mathbf{\Phi}=f(\mathbf{\Theta})$ that maps every system parameter choice $\mathbf{\Theta}\in \mathcal{P}$ to distinct GMD parameters $\mathbf{\Phi}$ to approach the respective SPDF. To further enable the parameter transform $\mathbf{\Phi}=f(\mathbf{\Theta})$ to automatically satisfy the normalization condition and other constraints of the GMD in Sec.~\ref{sec:3_1}, it is split into a deep neural network $\mathbf{y}=f_\text{NN}(\mathbf{\Theta})$ followed by a surjection $\mathbf{\Phi}=f_\text{SUR}(\mathbf{y})$ onto the parameter space of the GMD. Based on this two-step construction, the neural network $f_\text{NN}$ is capable of producing all and only all feasible density functions that the GMD can represent. Moreover, the normalization condition is implicitly encoded in the surjection such that the resulting unconstrained optimization of the neural network is simple.

\begin{figure}[!htb]
\center{\includegraphics[width=1\textwidth]
{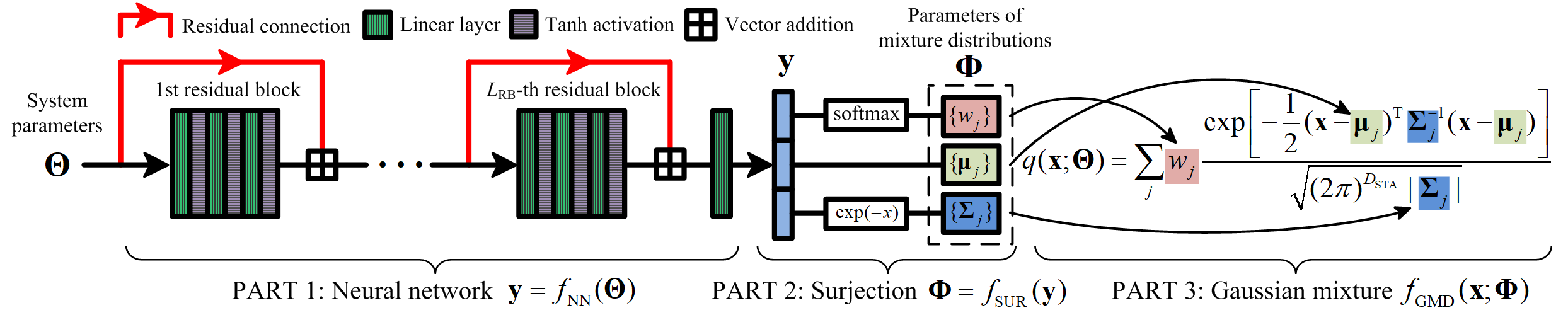}}
\caption{\label{fig:MDN} The architecture of the MDN $q(\mathbf{x};\mathbf{\Theta})=f_\text{GMD}(\mathbf{x};f_\text{SUR}(f_\text{NN}(\mathbf{\Theta})))$ for modelling the PAD that jointly solves the parameterized FPEs.}
\end{figure}

Figure~\ref{fig:MDN} shows the flowchart of the proposed MDN, which is detailed in Sec.~\ref{sec:3_1} and Sec.~\ref{sec:3_2}. With the MDN architecture, Sec.~\ref{sec:3_3} introduces a training algorithm to obtain the PAD that simultaneously solves the FPEs with any system parameters $\mathbf{\Theta}\in\mathcal{P}$.

\subsection{The Gaussian mixture distribution}\label{sec:gmm}
\label{sec:3_1}
A GMD $f_\text{GMD}(\mathbf{x};\mathbf{\Phi})$ is a convex combination of $N_{\text{GAU}}$ $D_\text{STA}$-D Gaussian distributions
\begin{equation}\label{eq:gaussian_mixture}
f_\text{GMD}(\mathbf{x};\mathbf{\Phi})=\sum_{j=1}^{N_\text{GAU}}w_j\phi(\mathbf{x};\bm{\mu}_j,\mathbf{\Sigma}_j),
\end{equation}
where the weights are in the convex set
\begin{equation}\label{eq:convex_set}
\mathcal{C}=\{[w_1,\cdots,w_{N_\text{GAU}}]^\top|\text{Every } w_j > 0 \text{ and } \sum_{j=1}^{N_\text{GAU}}w_j=1\},
\end{equation}
and $\phi(\mathbf{x};\bm{\mu}_j,\mathbf{\Sigma}_j)$ is the PDF of the $j$-th Gaussian component with the mean vector $\bm{\mu}_j=[\mu_{j1},\cdots,\mu_{jD_\text{STA}}]^\top$ and covariance matrix $\mathbf{\Sigma}_j$. Note that the positivity constraints of the weights $\{w_j\}_{j=1}^{N_{\text{GAU}}}$ correspond to the positivity condition of probability in Eq.~(\ref{eq:condition_nonega}) and the summation $\sum_{j=1}^{N_\text{GAU}}w_j=1$ corresponds to the normalization condition in Eq.~(\ref{eq:condition_norm}). The zero weights are excluded from Eq.~(\ref{eq:convex_set}) for the ease of the definition of the surjection in Sec.~\ref{sec:3_2}. The Gaussian components with weights close to zero are inactive as they have little influence on the summation. For simplification purpose we only consider diagonal covariance matrices, i.e., $\mathbf{\Sigma}_j=\text{diag}\{\sigma_{j1}^2,\cdots,\sigma_{jD_{\text{STA}}}^2\}$. Then, $\phi(\mathbf{x};\bm{\mu}_j,\mathbf{\Sigma}_j)$ is defined by
\begin{equation}\label{eq:gauss_distri}
\phi(\mathbf{x};\bm{\mu}_j,\mathbf{\Sigma}_j)=\frac{1}{\sqrt{(2\pi)^{D_{\text{STA}}}|\mathbf{\Sigma}_j|}}e^{-\frac{1}{2}(\mathbf{x}-\bm{\mu}_j)^\top\mathbf{\Sigma}_j^{-1}(\mathbf{x}-\bm{\mu}_j)}=\prod_{k=1}^{D_\text{STA}}\frac{1}{\sqrt{2\pi}\sigma_{jk}}e^{-\frac{(x_j-\mu_{jk})^2}{2\sigma_{jk}^2}}.
\end{equation}

The GMD defined in Eqs.~(\ref{eq:gaussian_mixture})-(\ref{eq:gauss_distri}) has $N_\text{GAU}(1+2D_\text{STA})$ parameters, denoted by
\begin{equation}
\begin{split}
\mathbf{\Phi}&=[\phi_1,\cdots,\phi_{N_\text{GAU}(1+2D_\text{STA})}]^\top\\
&=[w_1,\cdots,w_{N_\text{GAU}},\mu_{11},\cdots,\mu_{N_\text{GAU}D_\text{STA}},\sigma_{11},\cdots,\sigma_{N_\text{GAU}D_\text{STA}}]\in\mathcal{Q},
\end{split}
\end{equation}
including the $N_\text{GAU}$ weights $\{w_j\}_{j=1}^{N_\text{GAU}}$, the $N_\text{GAU}$ $D_\text{STA}$-D mean vectors $\{\bm{\mu}_j\}_{j=1}^{N_\text{GAU}}$, and the $N_\text{GAU}D_\text{STA}$ standard deviations $\{\sigma_{jk}\}_{j,k=1}^{N_\text{GAU}, D_\text{STA}}$, which are the diagonal elements of the $N_\text{GAU}$ $D_\text{STA}\times D_\text{STA}$ diagonal covariance matrices $\{\mathbf{\Sigma}_j\}_{j=1}^{N_\text{GAU}}$. The feasible GMD parameter domain is
\begin{equation}\label{eq:gmm_param_space}
\mathcal{Q}=\mathcal{C}\oplus \mathbb{R}^{N_\text{GAU}D_\text{STA}}\oplus\mathbb{R}_+^{N_\text{GAU}D_\text{STA}}\subset \mathbb{R}^{N_\text{GAU}(1+2D_\text{STA})},
\end{equation}
where $\oplus$ is the direct sum of spaces and $\mathbb{R}_+$ is the set of positive real numbers. 

Based on the modelling of the GMD, to solve the FPE is to find the best GMD parameters $\mathbf{\Phi}\in \mathcal{Q}$ such that $q(\mathbf{x};\mathbf{\Theta})=f_\text{GMD}(\mathbf{x};\mathbf{\Phi})$ approximates the SPDF $p(\mathbf{x};\mathbf{\Theta})$. Thus our learning goal is shifted to find an optimal parameter transform $f:\mathcal{P}\subset \mathbb{R}^{D_\text{PAR}}\rightarrow\mathcal{Q}$, which maps every $\mathbf{\Theta}$ in the system parameter domain $\mathcal{P}$ to the optimal $\mathbf{\Phi}=f(\mathbf{\Theta})$ in the GMD parameter domain $\mathcal{Q}$.

However, it is not straightforward to learn the transform $f:\mathcal{P}\rightarrow\mathcal{Q}$, as $\mathcal{Q}$ is a subspace of $\mathbb{R}^{N_\text{GAU}(1+2D_\text{STA})}$ with constraints. The first $N_\text{GAU}$ elements (the weights) are in the positive convex set $\mathcal{C}$ with the summation one and the last $N_\text{GAU}D_\text{STA}$ elements (the standard deviations) are positive. Deep learning is not good at directly learning these quantities. We should build a learning model that is capable of generating all possible GMD parameters to fully exploit the fitting capability of the GMD. At the same time, the optimization process should avoid explicitly enforcing the constraints of probability and the GMD. The two requirements are satisfied by the following construction.

\subsection{The deep learning-based parameter transform}\label{sec:dnn}
\label{sec:3_2}
We utilize a two-stage composite function to construct the parameter transform $\mathbf{\Phi}=f(\mathbf{\Theta})$ that maps the system parameters $\mathbf{\Theta}$ to the GMD parameters $\mathbf{\Phi}$
\begin{equation}\label{eq:param_map}
\left\{\begin{array}{l}
\mathbf{y}=f_\text{NN}(\mathbf{\Theta}) \in \mathbb{R}^{N_\text{GAU}(1+2D_\text{STA})},\\
\mathbf{\Phi}=f_\text{SUR}(\mathbf{y})\in \mathcal{Q}.
\end{array}\right.
\end{equation}
The first function $f_\text{NN}$ is a residual network (ResNet)~\cite{He2015DeepRL,Wang2024Deep,Feng2024Fusing}. It stacks a large number of linear layers to achieve an excellent nonlinear mapping capability and obtains an $N_\text{GAU}(1+2D_\text{STA})$-D vector $\mathbf{y}$ without any internal constraint. The constraints of the GMD parameters in Eq.~(\ref{eq:gmm_param_space}) are imposed and enclosed by the following surjective function $f_\text{SUR}:\mathbb{R}^{N_\text{GAU}(1+2D_\text{STA})}\rightarrow\mathcal{Q}$ that maps the intermediate vector $\mathbf{y}$ to the GMD parameters $\mathbf{\Phi}$. Hence, the fitting capability of the GMD $f_\text{GMD}$ is fully realized by the mapping capability of the ResNet $f_\text{NN}$, which is accomplished by our training algorithm in Sec.~\ref{sec:training}.

As shown in Fig.~\ref{fig:MDN}, the ResNet architecture is a traditional FCNN with residual connections such that information can flow among shallow and deep layers more quickly. The ResNet $f_\text{NN}$ cascades $L_\text{RB}$ residual blocks $g_k(\cdot)$ for $k=1,\cdots,L_\text{RB}$, followed by a final output linear layer $l_{3L_\text{RB}+1}(\cdot)$,
\begin{equation}\label{eq:f_nn}
\mathbf{y} = f_\text{NN}(\mathbf{\Theta})=l_{3L_\text{RB}+1}(g_{L_\text{RB}}(g_{L_\text{RB}-1}(\cdots g_2(g_1(\mathbf{\Theta}))))).
\end{equation}
A linear layer $l_k(\cdot)$ is defined by the affine transform
 \begin{equation}\label{eq:linear_layer}
l_k(\mathbf{z})=\mathbf{W}_k \mathbf{z} + \mathbf{b}_k,
\end{equation}
with the weight matrix $\mathbf{W}_k$ and bias vector $\mathbf{b}_k$. The $i$-th residual block $g_k(\cdot)$ is the vectorial summation of the residual connection $r_i(\cdot)$ and a three-layer neural network that cascades three linear layers and activation functions
\begin{equation}\label{eq:residual_block}
g_i(\mathbf{z})=r_i(\mathbf{z})\boxplus h(l_{3i}(h(l_{3i-1}(h(l_{3i-2}(\mathbf{z})))))), i=1,\cdots,L_\text{RB},
\end{equation}
where $\boxplus$ is the vectorial summation, and the elementwise hyperbolic tangent activation function $h(\cdot)$ is defined by
\begin{equation}
h(z)=\frac{e^z - e^{-z}} {e^z + e^{-z}}.
\end{equation}
We choose each of the $3L_\text{RB}$ linear layers in the $L_\text{RB}$ residual blocks has $N_\text{NEU}$ neurons~\cite{Wang2024Deep}. Only the first residual block has different input and output dimensions, as its input is the $D_\text{PAR}$-D system parameters $\mathbf{\Theta}$ while the output is an $N_\text{NEU}$-D vector. The input and output dimensions of all the other $L_\text{RB}-1$ residual blocks are $N_\text{GAU}$. The final linear layer $l_{3L_\text{RB}+1}$ in Eq.~(\ref{eq:f_nn}) maps the $N_\text{NEU}$-D output of the last residual block to the $N_\text{GAU}(1+2D_\text{STA})$-D vector $\mathbf{y}$.

In each residual block, the residual connection $r_i(\cdot)$ sends the input vector to the output side such that the forward signals and the backward derivatives bypass the three-layer neural network and the information flow can be accelerated. It is defined by
\begin{equation}
r_{i}(\mathbf{z})=\left\{\begin{array}{ll}
l_0(\mathbf{z}),&i=1,\\
\mathbf{z},&i=2,\cdots,L_\text{RB},
\end{array}\right.
\end{equation}
where $l_0(\cdot)$ is defined in Eq.~(\ref{eq:linear_layer}). In detail, starting from the second residual block, the residual connection is simply the identity function. In the first residual block, the input and output dimensions are different. An additional linear layer $l_0$ is required to adjust the dimension and ensure that the final vectorial summation in Eq.~(\ref{eq:residual_block}) can be carried out. The full parameters of the ResNet $f_\text{NN}$ consist of the weight matrices and the bias vectors $\{(\mathbf{W}_i,\mathbf{b}_i)\}_{i=0}^{3L_\text{RB}+1}$ of the $3L_\text{RB}+2$ linear layers.

The real $N_\text{GAU}(1+2D_\text{STA})$-D output $\mathbf{y}=[y_1,\cdots,y_{N_\text{GAU}(1+2D_\text{STA})}]^\top$ of the ResNet $f_\text{NN}$ is unconstrained. It is then mapped onto the parameter space $\mathcal{Q}$ of the GMD by the following surjective function $f_\text{SUR}$,
\begin{equation}\label{eq:surjection}
\begin{split}
\mathbf{\Phi}&=[\phi_1,\cdots,\phi_{N_\text{GAU}(1+2D_\text{STA})}]^\top=f_\text{SUR}(\mathbf{y}),\\
\phi_k&=\left\{\begin{array}{ll}
e^{y_k}/\sum_{k=1}^{N_\text{GAU}}e^{y_k},&1\leq k\leq N_\text{GAU},\\
y_k,&N_\text{GAU}+1\leq k \leq N_\text{GAU}(1+D_\text{STA}),\\
e^{-y_k},&N_\text{GAU}(1+D_\text{STA})+1\leq k \leq N_\text{GAU}(1+2D_\text{STA}),\\
\end{array}\right.
\end{split}
\end{equation}
The first part of $f_\text{SUR}$ is the softmax function, which establishes a surjective map from the first $N_\text{GAU}$ real elements of $\mathbf{y}$ onto the positive convex set $\mathcal{C}$ in Eq.~(\ref{eq:convex_set}), ensuring that the nonnegativity and normalization conditions are automatically satisfied. The second part takes the intermediate $N_\text{GAU}D_\text{STA}$ real elements of $\mathbf{y}$ as the $N_\text{GAU}$ unconstrained mean vectors of the GMD. The third part transforms the last $N_\text{GAU}D_\text{STA}$ real elements to the positive standard deviations of the GMD by the function $e^{-x}$. In other words, the last $N_\text{GAU}D_\text{STA}$ elements of $\mathbf{y}$ are the natural logarithms of the reciprocals of the standard deviations, which are real quantities. Eq.~(\ref{eq:surjection}) is a surjection from $\mathbb{R}^{N_\text{GAU}(1+2D_\text{STA})}$ to $\mathcal{Q}$, as every $\mathbf{y}\in\mathbb{R}^{N_\text{GAU}(1+2D_\text{STA})}$ corresponds to a GMD with the parameters $f_\text{SUR}(\mathbf{y})$ and for every parameter choice $\mathbf{\Phi}\in\mathcal{Q}$ of the GMD, the following $\mathbf{y}$ exists
\begin{equation}
\begin{split}
y_k&=\left\{\begin{array}{ll}
\ln \phi_k + \lambda,&1\leq k\leq N_\text{GAU},\\
\phi_k,&N_\text{GAU}+1\leq k \leq N_\text{GAU}(1+D_\text{STA}),\\
-\ln \phi_k,&N_\text{GAU}(1+D_\text{STA})+1\leq k \leq N_\text{GAU}(1+2D_\text{STA}),\\
\end{array}\right.
\end{split}
\end{equation}
for any arbitrary constant $\lambda$, 
such that $f_\text{SUR}(\mathbf{y})=\mathbf{\Phi}$.

In summary, the proposed MDN for modelling the SPDFs $p(\mathbf{x};\mathbf{\Theta})$ of the parameterized FPEs is 
\begin{equation}\label{eq:MDN}
q(\mathbf{x};\mathbf{\Theta})=f_\text{GMD}(\mathbf{x};f_\text{SUR}(f_\text{NN}(\mathbf{\Theta}))).
\end{equation}
It can exclusively represent all possible positive convex combinations of $N_\text{GAU}$ Gaussian density functions, and be the PAD if the ResNet $\mathbf{y}=f_\text{NN}(\mathbf{\Theta})$ can be suitably optimized.

\subsection{Network training}\label{sec:training}
\label{sec:3_3}

To effectively learn the ResNet $f_\text{NN}$ such that the MDN $q(\mathbf{x};\mathbf{\Theta})$ in Eq.~(\ref{eq:MDN}) fits the true SPDFs on all states $\mathbf{x}\in\mathcal{S}$ in Eq.~(\ref{eq:state_domain}) and all system parameters $\mathbf{\Theta}\in\mathcal{P}$ in Eq.~(\ref{eq:param_domain}), these two domains should be defined properly. The parameter domain $\mathcal{P}$ should be defined first such that the SPDF exists for each $\mathbf{\Theta}\in\mathcal{P}$. Then a large enough state domain $\mathcal{S}$ should be chosen such that $\int_\mathcal{S} p(\mathbf{x};\mathbf{\Theta})d\mathbf{x}\approx 1$ for every $\mathbf{\Theta}\in\mathcal{P}$.

The loss function of the MDN includes the FPE loss and the optional normalization condition. The FPE loss $L_\text{FP}$ evaluates the approximation of the MDN to the FPE solutions. As the MDN is a real-valued function with the domain $\mathcal{S}\oplus\mathcal{P}$, a $(D_\text{STA}+D_\text{PAR})$-D box spanned by the system states and parameters, the ideal optimization goal is to minimize the expected $L_1$ loss of the FPE condition in Eq.~(\ref{eq:condition_fp}), i.e., $\mathbb{E}_{(\mathbf{x},\mathbf{\Theta})\in\mathcal{S}\oplus\mathcal{P}}|\mathcal{L}_\text{FP}q(\mathbf{x};\mathbf{\Theta})|$. We adopt the robust $L_1$ loss as it is less sensitive to large errors and outliers than the $L_2$ loss~\cite{Barron2019General,Menezes2021review}. Thus in each training batch we uniformly sample $N_\text{SYS}$ choices of system parameters $\{\mathbf{\Theta}_j\}_{j=1}^{N_\text{SYS}}$ in the parameter domain $\mathcal{P}$. For each $\mathbf{\Theta}_j$, $N_\text{STA}$ states $\{\mathbf{x}_{jk}\}_{k=1}^{N_\text{STA}}$ are further uniformly sampled from the state domain $\mathcal{S}$. The FPE loss $L_\text{FP}$ is defined by the average $L_1$ loss of the $N_\text{SYS}N_\text{STA}$ parameter-state pairs $\{(\mathbf{\Theta}_j,\mathbf{x}_{jk})\}_{j,k=1}^{N_\text{SYS},N_\text{STA}}$
\begin{equation}\label{eq:loss}
L_\text{FP}=\frac{1}{N_\text{SYS}N_\text{STA}}\sum_{j=1}^{N_\text{SYS}}\sum_{k=1}^{N_\text{STA}}|\mathcal{L}_\text{FP}q(\mathbf{x}_{jk};\mathbf{\Theta}_j)|.
\end{equation}

Depending on whether the integral domain is the full space $\mathbb{R}^{D_\text{STA}}$ or the compact state domain $\mathcal{S}$, there are two normalization conditions. The first part of Eq.~(\ref{eq:surjection}) ensures that the MDN structurally satisfies the full normalization condition on $\mathbb{R}^{D_\text{STA}}$, i.e., Eq.~(\ref{eq:condition_norm}). Nevertheless, we explicitly consider the restricted normalization condition on the compact state domain $\mathcal{S}$~\cite{Xu2020Solving}, quantified by the following normalization loss $L_\text{NORM}$,
\begin{equation}\label{eq:loss_norm}
L_\text{NORM}=\frac{1}{N_\text{SYS}}\sum_{j=1}^{N_\text{SYS}}\left[\sum_{\mathbf{x} \in \mathcal{G}_\text{NORM}} q(\mathbf{x};\mathbf{\Theta}_j)\Delta \mathbf{x} - 1\right]^2,
\end{equation}
where $\mathcal{G}_\text{NORM}$ consists of the sample states on a regular grid that covers $\mathcal{S}$ and $\Delta \mathbf{x}$ is the $D_\text{STA}$-D differential element $\Delta \mathbf{x}$. The normalization loss $L_\text{NORM}$ is the average square error of the numerical integral of Eq.~(\ref{eq:condition_norm}) restricted to the state domain $\mathcal{S}$ for different system parameters. In the majority of our numerical experiments, i.e., Sec.~\ref{sec:sys_1d} to Sec.~\ref{sec:ex_6d_sys}, the full normalization condition coincides with the restricted one. Therefore, $L_\text{NORM}$ is redundant and can be removed from the loss function. However, without the explicit term $L_\text{NORM}$, sometimes the MDN only results in almost zero solutions in the state domain $\mathcal{S}$, e.g., Sec.~\ref{sec:ex_toggle_switch}. In such situation, the combined loss function $L_\text{FP}+L_\text{NORM}$ guarantees the PAD can be successfully learned and zero solutions on $\mathcal{S}$ are avoided. 

The natural boundary condition is also unnecessary in the loss function. By using the loss function $L_\text{FP}$ or $L_\text{FP}+L_\text{NORM}$, the MDN $q(\mathbf{x};\mathbf{\Theta})$ implicitly or explicitly satisfies the restricted normalization condition on the state domain $\mathcal{S}$, respectively. Furthermore, the FPE loss $L_\text{FP}$ ensures that the MDN has to approximate the true SPDF $p(\mathbf{x};\mathbf{\Theta})$ with the restricted normalization condition on $\mathcal{S}$. As a consequence, all the significant Gaussian components of the MDN are confined in $\mathcal{S}$ to meet the restricted normalization condition. The exponentially decaying characteristic of the Gaussian distribution thus guarantees that the probability density vanishes outside the boundary $\partial\mathcal{S}$.

The MDN in Eq.~(\ref{eq:MDN}) is a continuous function in both arguments $\mathbf{x}\in\mathcal{S}$ and $\mathbf{\Theta}\in\mathcal{P}$ as it is a composition of continuous functions. We then use automatic differentiation implemented in PyTorch library~\cite{pytorch2019} to calculate the derivatives in Eq.~(\ref{eq:condition_fp}) and resort to the ADAM~\cite{kingma2017adam} algorithm to optimize the ResNet. The full training algorithm is summarized in Algorithm~\ref{alg:1}. The main loop of Algorithm~\ref{alg:1} is to randomly sample the box $\mathcal{S}\oplus\mathcal{P}$ and optimize the MDN to fulfil the constraint of the FPE. For multi-dimensional systems with many parameters, the dimensionality of $\mathcal{S}\oplus\mathcal{P}$ is also very high, such that each training batch can only reach a sparse sampling. One can simply increase the number of training batches $K_\text{batch}$ to obtain denser samples and a better learning accuracy, thanks to our random sampling strategy in Algorithm~\ref{alg:1} for both states $\mathbf{x}$ and parameters $\mathbf{\Theta}$.

\begin{algorithm*}
\caption{Training the MDN-based PAD}\label{alg:1}
\begin{algorithmic}[1]
\REQUIRE The analytical FP operator $\mathcal{L}_\text{FP}$ Eq.~(\ref{eq:condition_fp}), the feasible state domain $\mathcal{S}$ and parameter domain $\mathcal{P}$, the number of training batches $K_\text{batch}$.
\STATE Initialize the ResNet $f_\text{NN}$.
\FOR {$r=1$ to $K_{\text{batch}}$}
	\STATE Uniformly sample $N_\text{SYS}$ $D_\text{PAR}$-D system parameters $\mathbf{\Theta}_j\in \mathcal{P}, j=1,\cdots,N_\text{SYS}$.
	\STATE Uniformly sample $N_\text{SYS}N_\text{STA}$ $D_\text{STA}$-D states $\mathbf{x}_{jk}\in \mathcal{S}, j=1,\cdots,N_\text{SYS}, k=1,\cdots,N_\text{STA}$.
	\STATE For each pair $(\mathbf{x}_{jk},\mathbf{\Theta}_j)$ evaluate the MDN output $q(\mathbf{x}_{jk};\mathbf{\Theta}_{j})$ in Eq.~(\ref{eq:MDN}) and the FP operator $\mathcal{L}_\text{FP}q(\mathbf{x}_{jk};\mathbf{\Theta}_{j})$ using automatic differentiation.
	\STATE Calculate the loss function Eq.~(\ref{eq:loss}) and use the ADAM algorithm to optimize the ResNet.
\ENDFOR
\RETURN The learnable parameters $\{(\mathbf{W}_i,\mathbf{b}_i)\}_{i=0}^{3L_\text{RB}+1}$ of the ResNet.
\end{algorithmic}
\end{algorithm*}

Compared to previous deep learning architectures for solving the FPE, the MDN has two key advantages in modelling the PAD.

(1) A flexible and concise Gaussian mixture fitting is adopted where the number of Gaussian components can be very small. Eq.~(\ref{eq:param_map}) implies that all the GMD parameters $\mathbf{\Phi}$, i.e., the combination weights, means, and diagonal covariance matrices, adapt to the parameters $\mathbf{\Theta}$ and are independent from the system states $\mathbf{x}$. Therefore, one can use a small number of Gaussian distributions, e.g., $N_\text{GAU}=50$ in the following numerical study, to simultaneously represent multi-dimensional multi-modal distributions governed by different system parameters $\mathbf{\Theta}\in\mathcal{P}$. The centers of the Gaussian components can move towards the regions with high probability and the covariance matrices also vary accordingly. The comparison in Fig.~\ref{fig:diff_num_gaussian} of Sec.~\ref{sec:discussion} shows that $N_\text{GAU}=10$ Gaussian components are enough to learn the SPDFs of 4D systems with 12 parameters. In contrast, the RBFNN uses thousands~\cite{Wang2023On,Xiao2024efficient} and hundreds~\cite{Wang2024Separable} of Gaussian components, respectively, for solving a single parameter choice. Because the standard deviations of the Gaussian distributions are fixed and the mean values are confined on a regular grid. Consequently, the number of Gaussian components increases considerably with the dimension of the system. 

(2) A simple loss function focuses solely on the FPE condition in Eq.~(\ref{eq:condition_fp}), making the optimization process efficient. It is due to that the nonnegativity and normalization conditions of probability, as well as the natural boundary condition, are encoded in the mapping structure, such that they are automatically satisfied. Therefore, these conditions are removed from the loss function. In contrast, the loss function of the FCNN~\cite{Xu2020Solving, Zhang2023Deep, Zhang2022Solving} explicitly includes the vital boundary condition and normalization condition, which require huge sample points on a regular grid or maintained by specific data structures.

Consequently, the MDN is grid-free in its construction and the optimization process. By optimizing the MDN, the parameterized FPEs are jointly solved. The resulting MDN can be the PAD to efficiently generate SPDFs with any $\mathbf{\Theta}\in \mathcal{P}$.

\section{Numerical experiments}
\label{sec:num_ex}
We now show the learning performance of the MDN-based PAD on several paradigmatic stochastic systems. In each case, a single MDN $q(\mathbf{x};\mathbf{\Theta})$ in Eq.~(\ref{eq:MDN}) is trained to simultaneously learn the SPDFs $p(\mathbf{x};\mathbf{\Theta})$ on all system parameters $\mathbf{\Theta}$ in the parameter domain $\mathcal{P}$ and all states $\mathbf{x}$ in the state domain $\mathcal{S}$. In all the experiments, we use the same network size, except that each system has specific dimensionality of the states and parameters. All the MDNs have $L_\text{RB}=6$ residual blocks, $N_\text{NEU}=50$ neurons in each linear layer and $N_\text{GAU}=50$ Gaussian components. Our method is implemented in PyTorch library~\cite{pytorch2019} and can be easily adapted to different systems by only modifying the specific dimensions, parameter ranges and the FP operators. The training and test are performed on a desktop with a GeForce RTX 4090 GPU and an Intel Core i9-13900K CPU.

\subsection{A 1-D tristable system with 7 parameters}
\label{sec:sys_1d}
We first consider the tristable system~\cite{Ma2020Precursor}
\begin{equation}\label{eq:sys_1d}
\dot{x}=V(x) + \sigma\xi,
\end{equation}
where $V(x)=ax^5+bx^4+cx^3+dx^2+ex+f$, $\xi$ is a SWGN and $\sigma$ is the noise intensity. The 1D state domain is $\mathcal{S}=\{x|x\in[-5, 5]\}$ and the 7D parameter domain is $\mathcal{P}=\{a\in [-2.5, -0.5]; b, c, d, e, f\in [-1, 1]; \sigma\in[0.2, 2.2]\}$. Here, the parameter $a$ is negative such that the SPDF exists. The state domain $\mathcal{S}$ is large enough such that for all system parameters $\mathbf{\Theta}\in\mathcal{P}$, the probability of the SPDF $p(x;\mathbf{\Theta})$ outside $\mathcal{S}$ is neglectable. The FP operator of $p(x)$ is
\begin{equation}
\mathcal{L}_\text{FP}p=-\frac{\partial}{\partial x}\left(Vp\right)+\frac{\sigma^2}{2}\frac{\partial^2p}{\partial x^2},
\end{equation}
where the dependent system parameters $\mathbf{\Theta}$ are omitted for the purpose of simplicity. The analytical SPDF is
\begin{equation}\label{eq:sys_1d_true_spdf}
p(x)\propto \text{exp}\left\{\frac{2}{\sigma^2}\int V(x)dx\right\},
\end{equation}
where the proportional symbol $\propto$ indicates that a constant is omitted from the right-hand side of the equation. When we calculate the true absolute or conditional SPDF on a regular grid or slice of states in $\mathcal{S}$, the constant can be determined by numerically calculating the normalization condition in Eq.~(\ref{eq:condition_norm}).

\begin{figure}[!htb]
\center{\includegraphics[width=1\textwidth]
{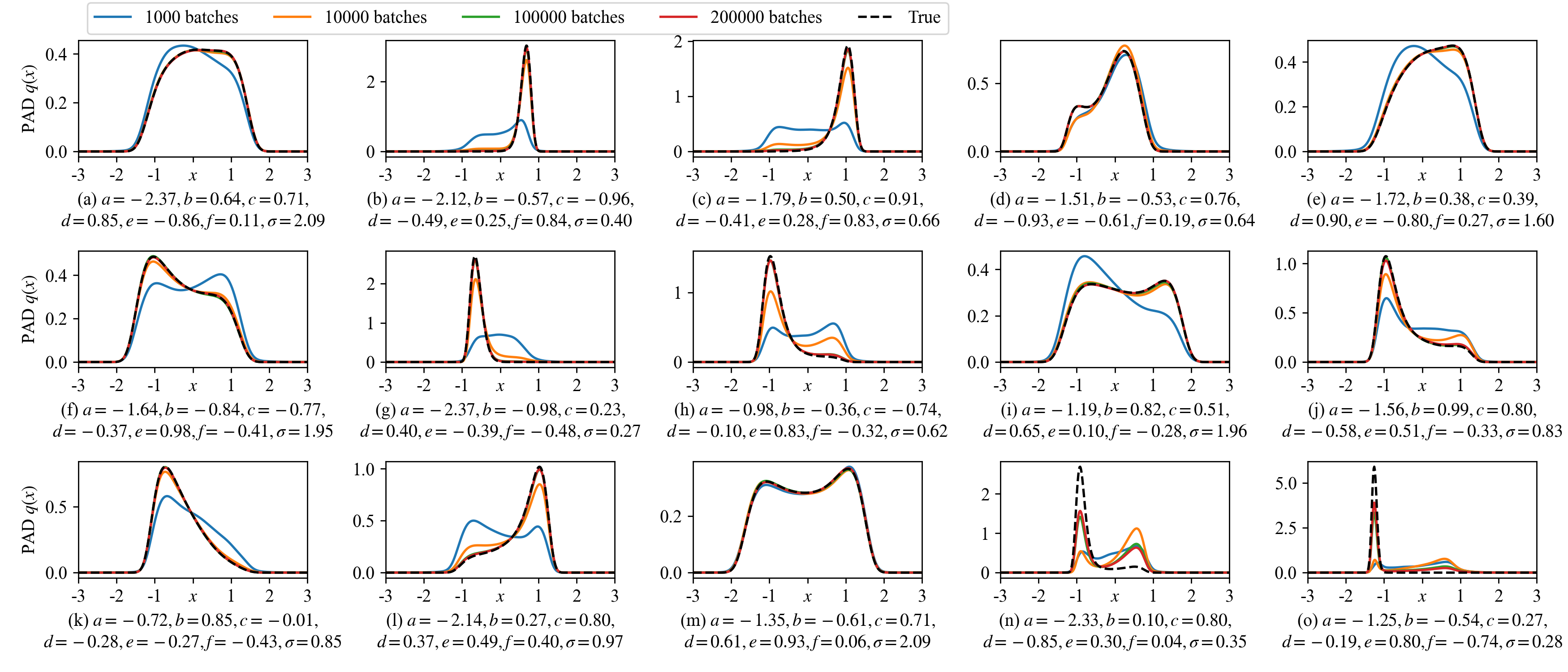}}
\caption{\label{fig:tristable} The learned SPDFs of the tristable system Eq.~(\ref{eq:sys_1d}) by the MDN-based PAD at different training batches. The parameters of the 15 systems are randomly sampled in the parameter domain $\mathcal{P}$. The true distributions are calculated from Eq.~(\ref{eq:sys_1d_true_spdf}). In the majority of cases, the learned solutions at 100,000 and 20,0000 training batches overlap, indicating the convergence.}
\end{figure}

A MDN-based PAD is trained for 200,000 batches where each batch evaluates $N_\text{SYS}\times N_\text{STA}=800\times 800=640,000$ random parameter-state pairs. Figure~\ref{fig:tristable} shows 15 evolved SPDFs at four different training batches. In the majority of cases, the learning converges before 100,000 batches and the SPDFs $p(x;\mathbf{\Theta})$ are well approximated by the PAD $q(x;\mathbf{\Theta})$. As designed in Eq.~(\ref{eq:surjection}), the normalization condition in Eq.~(\ref{eq:condition_norm}) is always satisfied. Though in the cases (n) and (o), 200,000 batches are still insufficient for the convergence, the learned solutions move towards the true SPDFs.

\begin{figure}[!htb]
\center{\includegraphics[width=1\textwidth]
{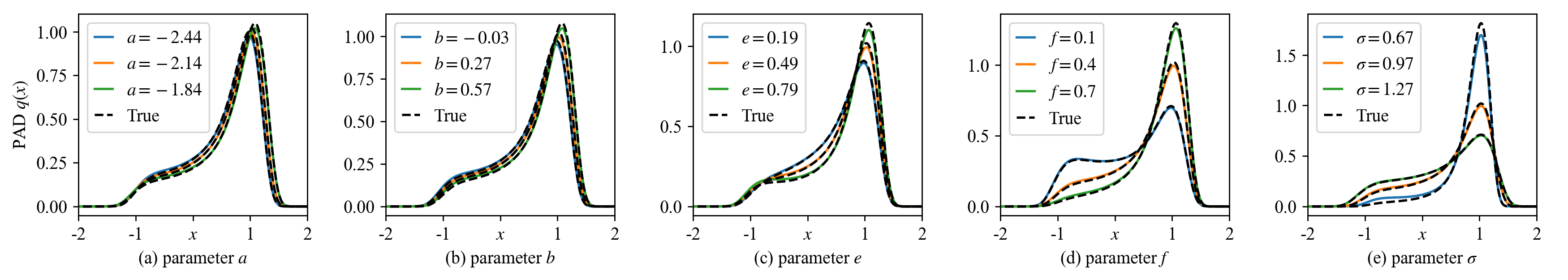}}
\caption{\label{fig:tristable2} The learned SPDFs with five groups of system parameters by the MDN-based PAD. In each group, one of $a=-2.14$, $b=0.27$, $e=0.49$, $f=0.4$, $\sigma=0.97$ changes, where the baseline parameter choice corresponds to Fig.~\ref{fig:tristable}(l).}
\end{figure}

In Fig.~\ref{fig:tristable2}, a baseline choice of system parameters are considered and the drift parameters $a$, $b$, $e$, $f$, and the noise intensity $\sigma$ are varied one by one. One can clearly see that the PAD always produces accurate SPDFs. Note that once the training is finished, the MDN can efficiently produce SPDFs by parallel computation on GPU. In our simulation, if each SPDF is numerically evaluated at 1000 states in the state domain $\mathcal{S}$, then the MDN can generate 100,000 SPDFs with different system parameters $\mathbf{\Theta}\in \mathcal{P}$ in just one second. Its high speed would significantly accelerate the subsequent analysis, such as the bifurcation study of noise induced transition.

\subsection{A 2-D Van der Pol system with 2 parameters}\label{sec:ex_van_der_pol}
The Van der Pol oscillator~\cite{Zhang2022Solving} is defined by
\begin{equation}\label{eq:sys_van_der_pol}
\left\{
\begin{array}{l}
\dot{x}=y,\\
\dot{y}=-\eta(x^2+y^2-1)y - x + \sigma \xi,
\end{array}
\right.
\end{equation}
where $\xi$ is a SWGN and $\sigma$ is the noise intensity. The 2D state domain is $\mathcal{S}=\{\mathbf{x}=[x,y]^\top|x,y\in[-5, 5]\}$ and the 2D parameter domain is $\mathcal{P}=\{\eta,\sigma\in[0.2, 1.0]\}$. The FP operator of $p(x,y)$ is
\begin{equation}
\mathcal{L}_\text{FP}p=-\frac{\partial}{\partial x}(yp)-\frac{\partial}{\partial y}\left[(-\eta(x^2+y^2-1)y - x)p\right]+\frac{\sigma^2}{2}\frac{\partial^2p}{\partial y^2},
\end{equation}
and the analytical SPDF is
\begin{equation}\label{eq:sys_van_der_pol_true_spdf}
p(x,y)\propto \text{exp}\left\{\frac{\eta}{\sigma^2}[x^2+y^2-\frac{1}{2}(x^2+y^2)^2]\right\}.
\end{equation}

\begin{figure}[!htb]
\center{\includegraphics[width=1\textwidth]
{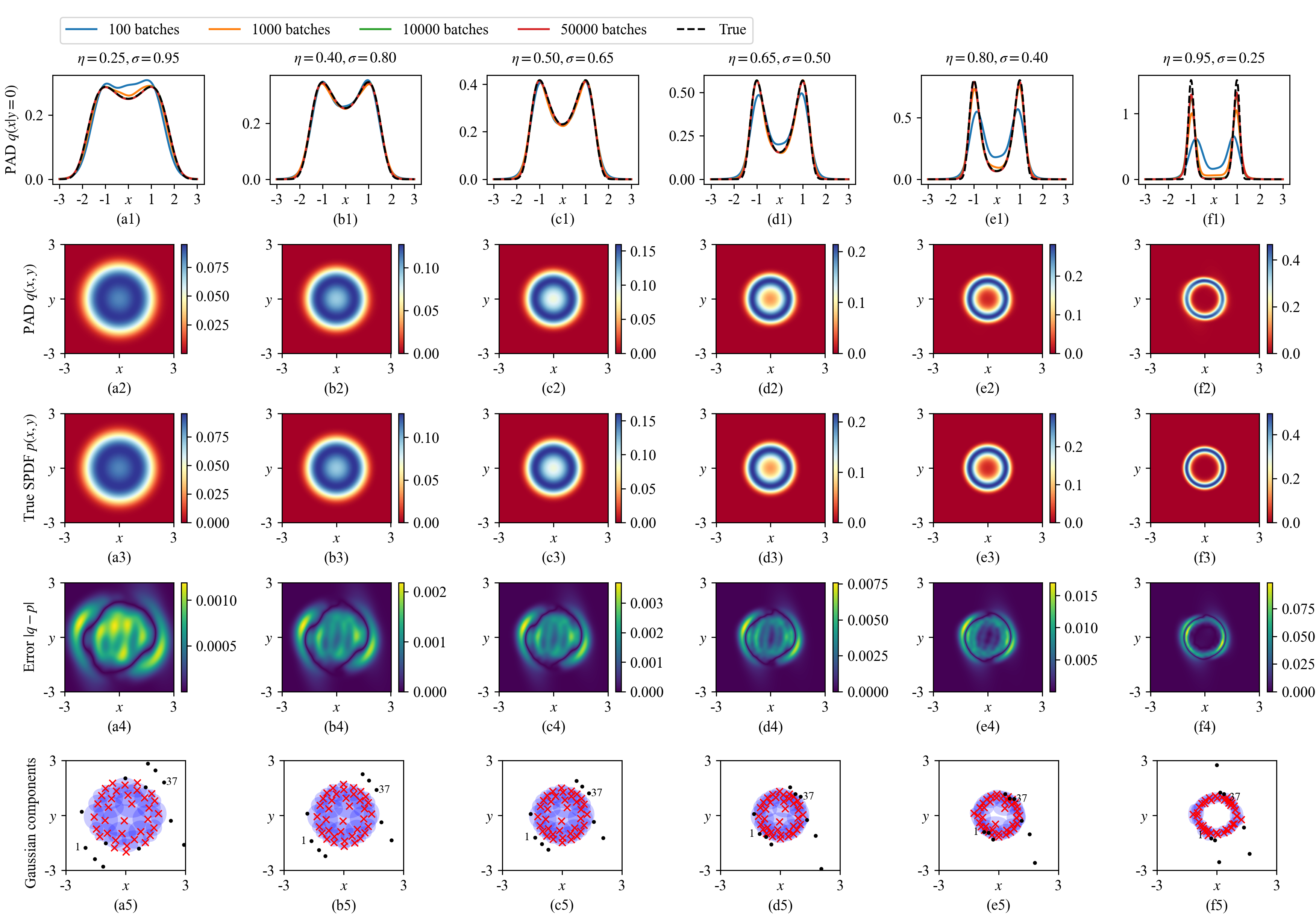}}
\caption{\label{fig:ex_van_der_pol} 6 groups of SPDFs of the Van der Pol system Eq.~(\ref{eq:sys_van_der_pol}) with different system parameters. The first row shows the learned distributions by the MDN-based PAD at different training batches. The second row to the fourth row draw the learned distributions at 50,000 training batches, the true distributions by Eq.~(\ref{eq:sys_van_der_pol_true_spdf}), and their differences, respectively. The last row details the adaptation of the $N_\text{GAU}=50$ Gaussian components. For the ease of visualization, only the 1st and 37th components are labelled. The means of active and inactive Gaussian components are plotted in red crosses and black dots, respectively. The blue ellipses indicate the high-probability regions defined in Eq.~(\ref{eq:high_prob_region}).}
\end{figure}

A MDN-based PAD is trained for 50,000 batches where each batch includes $N_\text{SYS}\times N_\text{STA}=800\times 800=640,000$ random parameter-state pairs. Figures~\ref{fig:ex_van_der_pol}(a1)-(a6) shows the learned conditional distributions $q(x|y=0)$ of 6 parameter choices. The MDN convergences quickly as the distributions are nearly convergent at 1000 training batches. The second and third rows of Fig.~\ref{fig:ex_van_der_pol} show the learned SPDF $q(x,y)$ by the PAD and the true distributions by Eq.~(\ref{eq:sys_van_der_pol_true_spdf}) and their differences are demonstrated in the fourth row of Fig.~\ref{fig:ex_van_der_pol}. One can clearly see that the PAD produces decent approximations. 

A core merit of the MDN-based PAD is that the $N_\text{GAU}=50$ Gaussian distributions derived from a single ResNet can adapt to different system parameters. To visualize this feature, we consider that in the view of fitting multi-dimensional functions, the GMD in Eq.~(\ref{eq:gaussian_mixture}) is a summation of $N_\text{GAU}$ Gaussian-shaped functions $\{w_j\phi(\cdot;\bm{\mu}_j,\mathbf{\Sigma}_j)\}_{j=1}^{N_\text{GAU}}$. Thus we define the $j$-th Gaussian component is active if the following high-probability region is nonempty,
\begin{equation}\label{eq:high_prob_region}
H_j = \{\mathbf{x}|w_j\phi(\mathbf{x};\bm{\mu}_j,\mathbf{\Sigma}_j)>0.005\},
\end{equation}
otherwise the $j$-th component is inactive. In the last row of Fig.~\ref{fig:ex_van_der_pol}, the active and inactive Gaussian components are plotted separately. The majority of active Gaussian components are gathered around the high-probability regions of the respective SPDFs, while a few components are inactive alternately. For example, the 1st and 37th components are inactive and distant from the high-probability region in Fig.~\ref{fig:ex_van_der_pol}(a5), while they move to the high-probability region in Fig.~\ref{fig:ex_van_der_pol}(f5). Therefore, the MDN can adaptively adjust a very limited number of Gaussian distributions to learn multi-dimensional systems with multiple parameters.

\subsection{A 4-D system with 12 parameters}\label{sec:ex_4d_sys}
Consider the following 4D system~\cite{Zhang2022Solving}
\begin{equation}\label{eq:4d_sys}
\left\{
\begin{array}{l}
\dot{x}_1=y_1,\\
\dot{y}_1=-ay_1-\frac{1}{M}\frac{\partial V(x_1,x_2)}{\partial x_1} + \sigma_1 \xi_1,\\
\dot{x}_2=y_2,\\
\dot{y}_2=-by_2-\frac{1}{I}\frac{\partial V(x_1,x_2)}{\partial x_2} + \sigma_2 \xi_2,\\
\end{array}
\right.
\end{equation}
where $V(x_1,x_2)=k_1x_1^2+k_2x_2^2+\varepsilon(\lambda_1x_1^4+\lambda_2x_2^4+\mu x_1^2x_2^2)$, $\xi_1$ and $\xi_2$ are two independent SWGNs, and $\sigma_1$ and $\sigma_2$ are the noise intensities, respectively. The 4D state domain is $\mathcal{S}=\{\mathbf{x}=[x_1,x_2,y_1,y_2]^\top|x_1,x_2,y_1,y_2\in[-10, 10]\}$ and the 12D parameter domain is $\mathcal{P}=\{a, b\in[0.2, 1.2];k_1,k_2\in[-1, 1]; \lambda_1,\lambda_2,\mu\in [0.1, 0.5];\varepsilon,M,I \in[0.5, 2];\sigma_1, \sigma_2\in[1, 2.5]\}$. The FP operator of $p(x_1, x_2, y_1, y_2)$ is
\begin{equation}
\mathcal{L}_\text{FP}p=-\frac{\partial (y_1 p)}{\partial x_1}-\frac{\partial (y_2 p)}{\partial x_2}
-\frac{\partial\left[(-ay_1-\frac{1}{M}\frac{\partial V}{\partial x_1})p\right]}{\partial y_1}
-\frac{\partial\left[(-by_2-\frac{1}{I}\frac{\partial V}{\partial x_2})p\right]}{\partial y_2}
+\frac{\sigma_1^2}{2}\frac{\partial^2 p}{\partial y_1^2}
+\frac{\sigma_2^2}{2}\frac{\partial^2 p}{\partial y_2^2}.
\end{equation}
When
\begin{equation}\label{eq:ana_4d}
\sigma_1^2M/a=\sigma_2^2I/b=2T,
\end{equation}
the analytical SPDF is expressed by
\begin{equation}\label{eq:4d_sys_true_spdf}
p(x_1,x_2,y_1,y_2)\propto \text{exp}\left\{-\frac{1}{T}V(x_1,x_2)-\frac{a}{\sigma_1^2}y_1^2-\frac{b}{\sigma_2^2}y_2^2\right\},
\end{equation}
otherwise the SPDF has to be solved numerically. It is worth noting that the 12 parameters are redundant since the four parameters $\varepsilon$, $\lambda_1$, $\lambda_2$, and $\mu$ in the function $V$ have only three degrees of freedom. The MDN can nicely handle such situation. In Algorithm~\ref{alg:1}, random samples in the parameter domain $\mathcal{P}$ sufficiently explore the respective variations of redundant parameters.

\begin{figure}[!htb]
\center{\includegraphics[width=1\textwidth]
{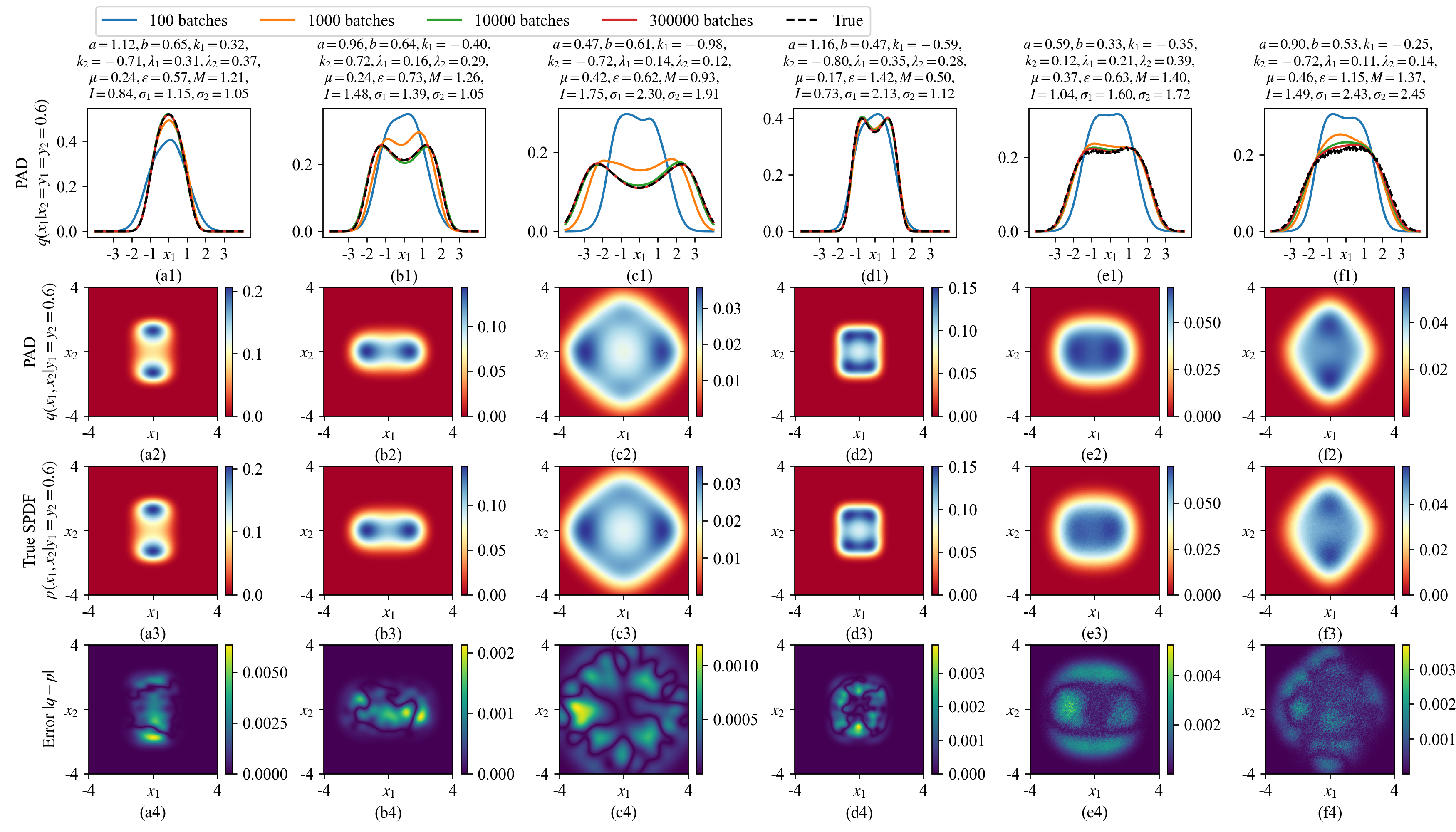}}
\caption{\label{fig:ex_4d_sys} 6 groups of SPDFs of the 4D system Eq.~(\ref{eq:4d_sys}) with different system parameters. The first row shows the learned 1D conditional distributions $q(x_1|x_2=y_1=y_2=0.6)$ by the MDN-based PAD at different training batches. The second row to the fourth row draw the learned 2D conditional distributions $q(x_1,x_2|y_1=y_2=0.6)$ at 300,000 training batches, the true distributions, and their differences, respectively. The true distributions in (a3)-(d3) are calculated by Eq.~(\ref{eq:4d_sys_true_spdf}) as the system parameters satisfy the condition Eq.~(\ref{eq:ana_4d}) where the analytical SPDF exists. As the last two cases have no analytical SPDF, the true distributions in (e3) and (f3) are calculated by Monte-Carlo simulation detailed in Appendix~\ref{sec:app_true_SPDF}.}
\end{figure}

We train a MDN-based PAD for 300,000 batches, where each batch uses $N_\text{SYS}\times N_\text{STA}=500\times 500=250,000$ random parameter-state pairs. Figure~\ref{fig:ex_4d_sys} shows the learned solutions of 6 groups of different system parameters. In each case, the first row of Fig.~\ref{fig:ex_4d_sys} shows the learned conditional distributions $q(x_1|x_2=y_1=y_2=0.6)$. The second and third rows show the learned marginal distributions $q(x_1, x_2|y_1=y_2=0.6)$ and the true distributions, respectively, while their differences are plotted in the last row of Fig.~\ref{fig:ex_4d_sys}. All the SPDFs are well restored by a single PAD.

\subsection{A 6-D system with 9 parameters}\label{sec:ex_6d_sys}
Consider the following 6D system~\cite{Zhang2022Solving}
\begin{equation}\label{eq:6d_sys}
\left\{
\begin{array}{l}
\dot{x}_i=y_i,\\
\dot{y}_i=-k_i y_i-\frac{\partial V(x_1,x_2,x_3)}{\partial x_i} + \sigma_i \xi_i,\\
\end{array}
\right., i=1,2,3,
\end{equation}
where $V(x_1,x_2,x_3)=0.25x_1(x_2+x_3)+\sum_{i=1}^3 \lambda_i x_i^2$, $\{\xi_i\}_{i=1}^3$ and $\{\sigma_i\}_{i=1}^3$ are three independent SWGNs and the corresponding noise intensities, respectively. The 6D state domain is $\mathcal{S}=\{\mathbf{x}=[x_1,x_2,x_3,y_1,y_2,y_3]^\top|x_i,y_i\in[-8, 8], i=1,2,3\}$ and the 9D parameter domain is $\mathcal{P}=\{k_i, \lambda_i\in[0.5, 1.5];\sigma_i\in[0.5, 2],i=1,2,3\}$. The FP operator of $p(x_1, x_2,x_3, y_1, y_2, y_3)$ is
\begin{equation}
\mathcal{L}_\text{FP}p=\sum_{i=1}^3\left[-\frac{\partial (y_i p)}{\partial x_i}
-\frac{\partial\left((-k_iy_i-\frac{\partial V}{\partial x_i})p\right)}{\partial y_i}
+\frac{\sigma_i^2}{2}\frac{\partial^2 p}{\partial y_i^2}
\right].
\end{equation}
When 
\begin{equation}\label{eq:ana_6d}
k_1/\sigma_1^2=k_2/\sigma_2^2=k_3/\sigma_3^2=1/T,
\end{equation}
the analytical SPDF can be expressed as
\begin{equation}\label{eq:6d_sys_true_spdf}
p(x_1,x_2,x_3,y_1,y_2,y_3)\propto \text{exp}\left[-\frac{1}{T}(2V+y_1^2+y_2^2+y_3^2)\right].
\end{equation}

\begin{figure}[!htb]
\center{\includegraphics[width=1\textwidth]
{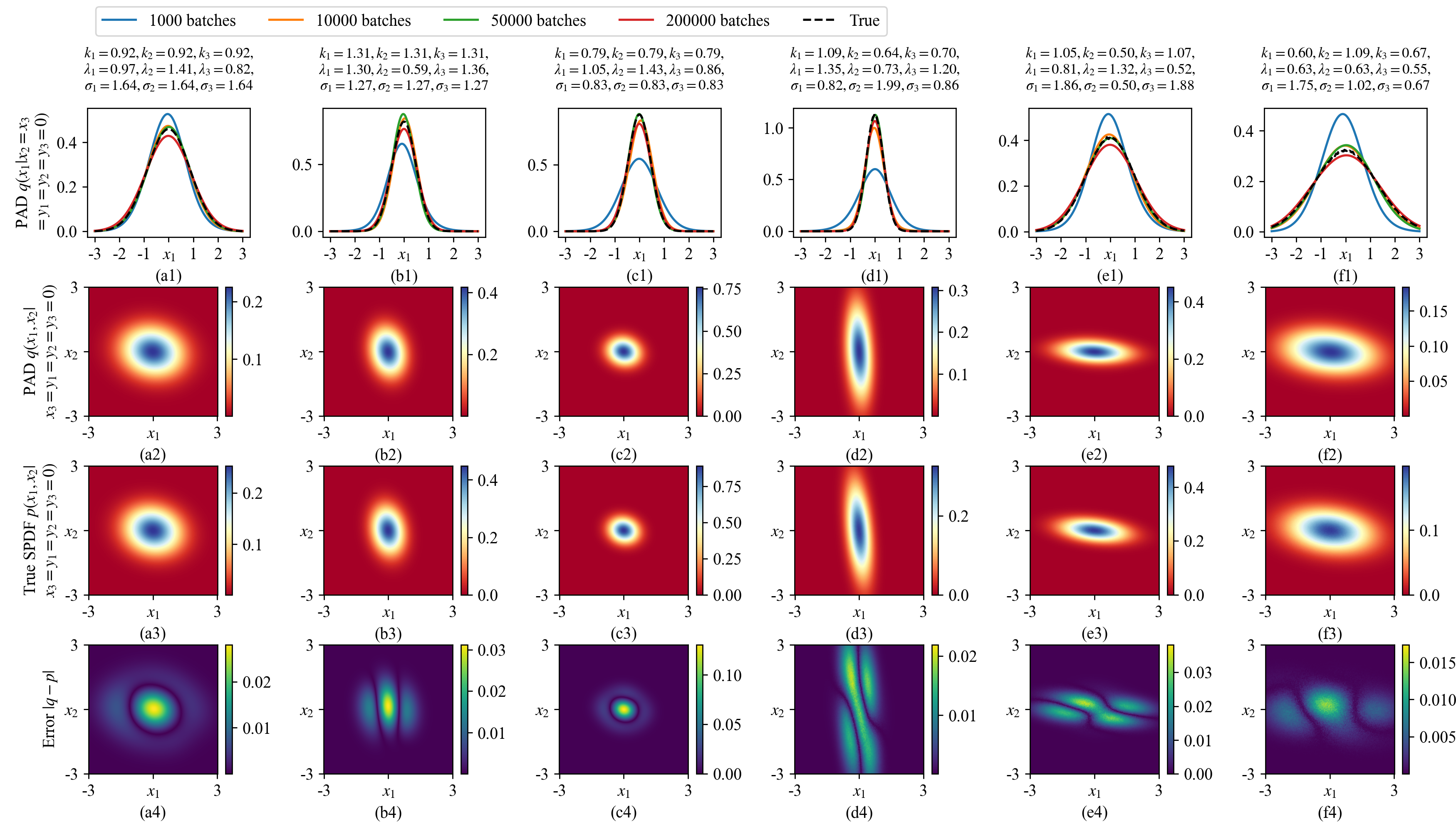}}
\caption{\label{fig:ex_6d_sys} 6 groups of learned SPDFs of the 6D system Eq.~(\ref{eq:6d_sys}) with different system parameters. The first row shows the learned 1D conditional distributions $q(x_1|x_2=x_3=y_1=y_2=y_3=0)$ by the MDN-based PAD at different training batches. The second and third rows draw the learned 2D conditional distributions $q(x_1,x_2|x_3=y_1=y_2=y_3=0)$ at 200,000 training batches and the true distributions, respectively, while their differences are detailed in the fourth row. The true SPDFs in (a3)-(c3) are calculated by Eq.~(\ref{eq:6d_sys_true_spdf}) as the analytical SPDFs exist. The true SPDFs of (d3)-(f3) are calculated by Monte-Carlo simulation detailed in Appendix~\ref{sec:app_true_SPDF}.}
\end{figure}

We train a MDN-based PAD for 200,000 batches where each batch includes $N_\text{SYS}\times N_\text{STA}=450\times 450=202,500$ random parameter-state pairs. Figure~\ref{fig:ex_6d_sys}(a1)-(f1) show six groups of conditional distributions $q(x_1|x_2=x_3=y_1=y_2=y_3=0)$ during the training process, from which we can see that the MDN produce good approximations. The second row, third row and fourth row show the learned conditional distributions $q(x_1,x_2|x_3=y_1=y_2=y_3=0)$ by the MDN, the true distributions and their differences, respectively, which indicate that all the distributions can be properly learned.

\subsection{A 2-D genetic toggle switch system with 5 parameters}\label{sec:ex_toggle_switch}
In all the previous examples the FPE loss $L_\text{FP}$ in Eq.~(\ref{eq:loss}) is used solely to train the MDN. In this section, we show a failure case to indicate its possible weakness and highlight the complexity of solving the FPE. Consider the genetic toggle switch system~\cite{Zhai2022deep,Peng2024deep}
\begin{equation}\label{eq:toggle_switch}
\left\{
\begin{array}{l}
\dot{x}=\frac{a+x^2}{a+x^2+y^2}-bx+\sigma_1\xi_1,\\
\dot{y}=\frac{a+y^2}{a+x^2+y^2}-cy+\sigma_2\xi_2,
\end{array}
\right.
\end{equation}
where $\xi_1$ and $\xi_2$ are two independent SWGNs and their intensities are $\sigma_1$ and $\sigma_2$, respectively. The 2D state domain is $\mathcal{S}=\{\mathbf{x}=[x,y]^\top|x,y\in[-0.5, 2]\}$ and the 5D parameter domain is $\mathcal{P}=\{a\in[0.1, 1];b,c\in[0.5,1.5 ];\sigma_1,\sigma_2\in[0.05, 0.3]\}$. The FP operator of $p(x,y)$ is
\begin{equation}
\mathcal{L}_\text{FP}p=-\frac{\partial}{\partial x}\left[\left(\frac{a+x^2}{a+x^2+y^2}-bx\right)p\right]-\frac{\partial}{\partial y}\left[\left(\frac{a+y^2}{a+x^2+y^2}-cy\right)p\right]+\frac{\sigma_1^2}{2}\frac{\partial^2 p}{\partial x^2}+\frac{\sigma_2^2}{2}\frac{\partial^2 p}{\partial y^2},
\end{equation}
without analytical stationary solutions.
\begin{figure}[!htb]
\center{\includegraphics[width=1\textwidth]
{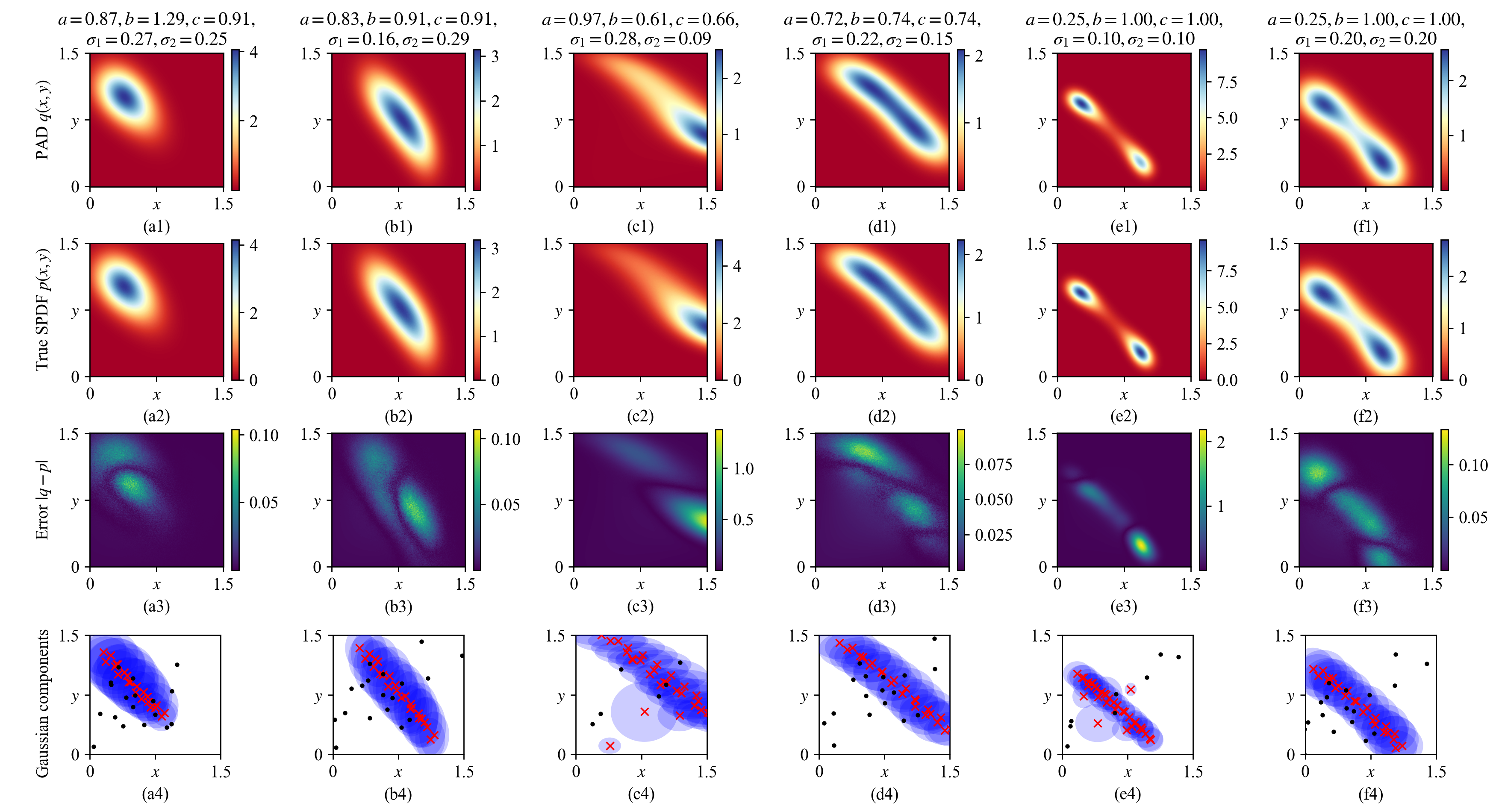}}
\caption{\label{fig:ex_toggle_switch} 6 groups of learned SPDFs of the toggle switch system Eq.~(\ref{eq:toggle_switch}) with different system parameters. The distributions in the restricted state domain $[0, 1.5]\times [0, 1.5]$ are shown. The first and second rows draw the learned distributions $q(x, y)$ by the MDN-based PAD at 100,000 training batches and the true distributions by Monte-Carlo simulation in Appendix~\ref{sec:app_true_SPDF}, respectively, while their differences are detailed in the third row. The last row details the mean values of the Gaussian components and the high-probability regions defined in Eq.~(\ref{eq:high_prob_region}).}
\end{figure}

Our numerical experiment shows that the single FPE loss $L_\text{FP}$ in Eq.~(\ref{eq:loss}) fails to learn the correct PAD. In the $N_\text{GAU}$ Gaussian components produced by the MDN, one component has a weight close to one and very large variances, while the others have weights close to zero. Their joint distribution is almost a zero solution in the compact state domain $\mathcal{S}$. Though the full normalization condition in Eq.~(\ref{eq:condition_norm}) is satisfied, the restricted normalization condition on $\mathcal{S}$ fails. The failure can be detected in the early stage of the optimization process in Algorithm~\ref{alg:1} by monitoring that the probability densities of the samples in $\mathcal{S}$ vanish quickly. However, this failure cannot be fixed by enlarging $\mathcal{S}$. Zero solutions in $\mathcal{S}$ may also happen to the previous systems in Sec.~\ref{sec:sys_1d} to Sec.~\ref{sec:ex_6d_sys} if the $L_2$ error is used in the FPE loss, signifying the crucial usage of the robust $L_1$ error.

To successfully train the MDN, the two-term loss function $L_\text{FP}+L_\text{NORM}$ is utilized, where the restricted normalization condition in Eq.~(\ref{eq:loss_norm}) is explicitly included to force the Gaussian components focusing on the compact state domain $\mathcal{S}$. We train a MDN-based PAD for 100,000 batches where each batch includes $N_\text{SYS}\times N_\text{STA}=500\times 500=250,000$ random parameter-state pairs to calculate the FPE loss Eq.~(\ref{eq:loss}). In addition, $\mathcal{G}_\text{NORM}$ includes another 2601 states on a $51\times 51$ regular grid of $\mathcal{S}$ to calculate the normalization loss Eq.~(\ref{eq:loss_norm}), with the area element $\Delta \mathbf{x}=(2.5/50)^2=0.0025$. Figure~\ref{fig:ex_toggle_switch} details six groups of the learned distributions $q(x,y)$ by the PAD, the true distributions $p(x,y)$ and their differences in the first, second, and third rows, respectively. Though all the distributions can be well learned in general, detail structure may not be perfectly restored. Comparing Figs.~\ref{fig:ex_toggle_switch}(e1) and (e2), the peak in the lower right region is not precisely reconstructed. Nevertheless, the last row of Fig.~\ref{fig:ex_toggle_switch} shows the Gaussian components properly adapt to the true SPDFs.

\section{Discussion}
\label{sec:discussion}

\begin{figure}[!htb]
\center{\includegraphics[width=1\textwidth]
{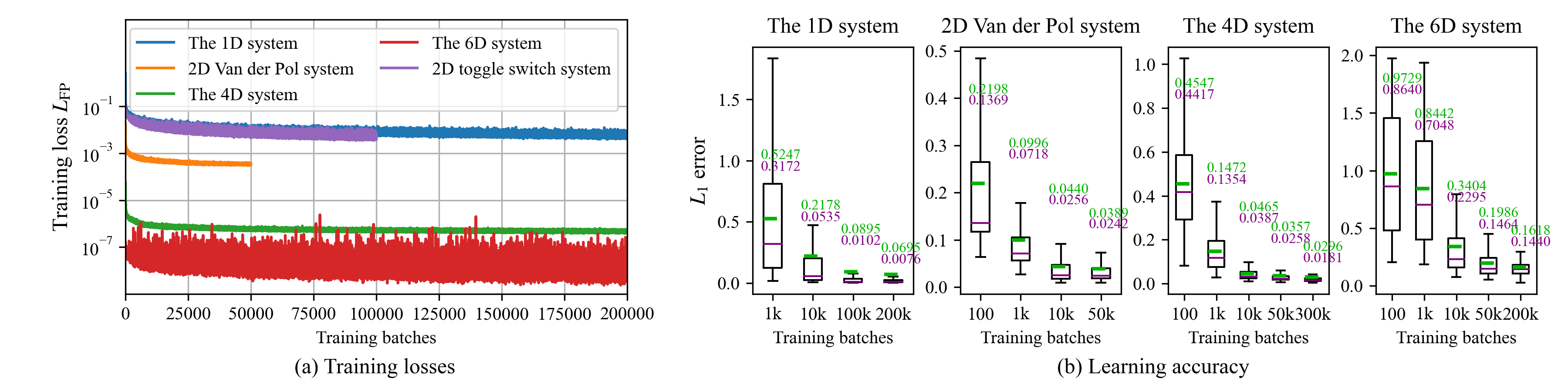}}
\caption{\label{fig:losses} The training FPE losses in Eq.~(\ref{eq:loss}) and test errors in Eq.~(\ref{eq:l1_error}) of the PADs in Sec.~\ref{sec:num_ex}. The test errors in (b) are summarized on 10,000 random choices of system parameters. The upper green and lower purple texts indicate the mean and median errors, respectively.}
\end{figure}

The training process of the MDN-based PAD is simple, as there are no sensible hyper-parameters. The training settings of the MDNs in Sec.~\ref{sec:num_ex} are listed in Tab.~\ref{tab:nn}. In all the experiments, we choose the numbers $N_\text{SYS}$ and $N_\text{STA}$ are equal. They should be chosen as large as possible to sufficiently utilize the GPU memory. Figure~\ref{fig:losses}(a) shows the FPE losses $L_\text{FP}$ during the training processes. When the training continues, the losses decrease steadily with some stochastic fluctuations due to that the $N_\text{SYS}N_\text{STA}$ parameter-state pairs in each batch are randomly sampled. It also implies that the training dataset is unlimited. Therefore, if we continue the training process, the learning accuracy increases monotonically, though the rate of improvement in accuracy tends to diminish.

\begin{table*}[!htb]
\scriptsize
\centering
\caption{\label{tab:nn}A comparison of the architectures and training processes of the MDN-based PADs in Sec.~\ref{sec:num_ex}.}
\begin{threeparttable}
\begin{tabular}{ccccccc}
\toprule
System & $D_\text{STA}$/$D_\text{PAR}$/  & \# Weights & Training & GPU &  Training time & Mean/median\\
&$N_\text{SYS}$/$N_\text{STA}$\tnote{*}& of MDN&batches&Memory&for 50,000 batches&error\tnote{**}\\
\midrule
The 1D system             & 1/7/800/800 & 51,800 &200,000&11,666 MiB &  1.3 hours  &0.0695/0.0076\\
2D Van der Pol & 2/2/800/800 & 56,400 &50,000&18,602 MiB & 1.65 hours  &0.0242/0.0389\\
The 4D system             & 4/12/800/800 & 67,600 &300,000&20,248 MiB & 2.7 hours  &0.0296/0.0181\\
The 6D system             & 6/9/450/450 & 77,500 &200,000&22,136 MiB & 3.4 hours  &0.1618/0.1440\\
2D toggle switch\tnote{***} & 2/5/500/500 & 56,700 &100,000&17,008 MiB& 1.8 hours&NA\\
\bottomrule
\end{tabular}
\begin{tablenotes}
\footnotesize
	\item[*] The dimension of the system, the number of system parameters, the number of groups of different system parameters in each training batch, and the number of random states used for every choice of system parameters.
	\item[**] The statistics are averaged on the $L_1$ errors in Eq.~(\ref{eq:l1_error}) between the PAD at the final training batch and the true SPDFs on 10,000 random parameter choices. See Appendix~\ref{sec:app_l1_dist} for the implementation detail.
	\item[***] Without the 2601 additional states for calculating the normalization loss, the training takes 1.24 hours and uses 9216 MiB GPU memory, though the obtained MDN fails to capture the SPDFs. We do not calculate the mean/median error as the system has no analytical solutions.
\end{tablenotes}
\end{threeparttable}
\end{table*}

To quantitatively evaluate the test error of the PAD, we introduce the following $L_1$ error between the true SPDF $p(\mathbf{x};\mathbf{\Theta})$ and the PAD $q(\mathbf{x};\mathbf{\Theta})$
\begin{equation}\label{eq:l1_error}
d_\text{L1}(p,q;\mathbf{\Theta})=\int_{\mathcal{S}}|p(\mathbf{x};\mathbf{\Theta})-q(\mathbf{x};\mathbf{\Theta})|d\mathbf{x}.
\end{equation}
We further define that the error metric of the PAD on systems with the parameter domain $\mathcal{P}$ is the average $L_1$ error of 10,000 different parameters $\{\mathbf{\Theta}_j\}_{j=1}^{10,000}$ randomly sampled in $\mathcal{P}$ with available analytical SPDFs. The numerical calculation of Eq.~(\ref{eq:l1_error}) is detailed in Appendix~\ref{sec:app_l1_dist}. The test errors of the PADs in Secs.~\ref{sec:sys_1d} to \ref{sec:ex_6d_sys} are listed in the last column of Tab.~\ref{tab:nn} and plotted in Fig.~\ref{fig:losses}(b). In every case, the average test error of the whole parameter domain decreases steadily when the training continues.

\begin{figure}[!htb]
\center{\includegraphics[width=1\textwidth]
{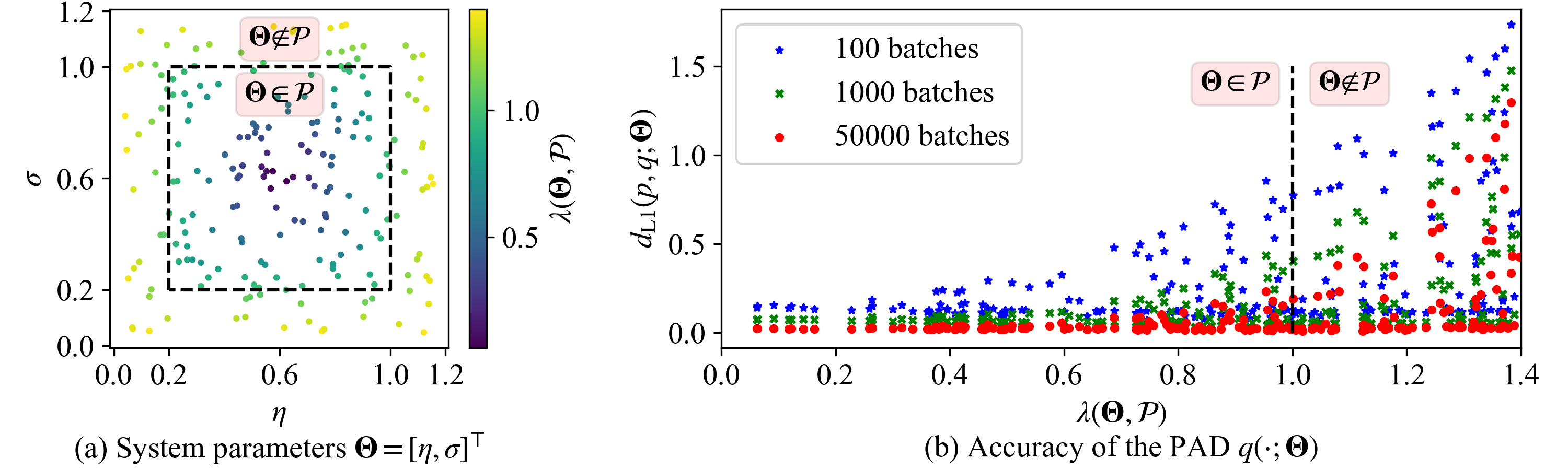}}
\caption{\label{fig:param_vs_error} The accuracy and the extrapolation ability of the MDN-based PAD during the training process of the Van der Pol system in Sec.~\ref{sec:ex_van_der_pol}. (a) 200 system parameters $\mathbf{\Theta}$ are uniformly sampled in the augmented parameter domain where the ratio $\lambda$ indicates how close the parameter to the center of the training parameter domain $\mathcal{P}$. (b) The $L_1$ errors between the PAD $q(\mathbf{x};\mathbf{\Theta})$ at the 100, 1000, 50,000 training batches and the true solutions $p(\mathbf{x};\mathbf{\Theta})$.}
\end{figure}

To investigate the extrapolation ability of the PAD, the correlation between its test error and the location of the parameter choice in the parameter domain is considered. We adopt the terms in Eq.~(\ref{eq:param_domain}) and define the following distance ratio
\begin{equation}
\lambda(\mathbf{\Theta}, \mathcal{P}) = \max_{1\leq i\leq D_\text{PAR}} \frac{2\left|\theta_i - \frac{\theta_i^\text{min}+\theta_i^\text{max}}{2}\right|}{\theta_i^\text{max}-\theta_i^\text{min}},
\end{equation}
which measures how close the parameters $\mathbf{\Theta}$ to the center of the parameter domain $\mathcal{P}$. $\lambda=0$ means that $\mathbf{\Theta}$ is at the center of $\mathcal{P}$ and $\lambda$ increases when $\mathbf{\Theta}$ moves towards the boundary of $\mathcal{P}$, where $\lambda=1$. If $\lambda>1$, the parameters are outside $\mathcal{P}$, indicating the parameter choice is not included in the training domain $\mathcal{P}$. As shown in Fig.~\ref{fig:param_vs_error}(a), we uniformly sample 200 system parameters of the Van der Pol system in the 2D domain with $\lambda\in[0, 1.4]$, a box bigger than $\mathcal{P}$. The $L_1$ errors of the PAD at three training stages are plotted in Fig.~\ref{fig:param_vs_error}(b). One can clearly see that more training batches lead to lower errors. After 50,000 training batches, the $L_1$ errors between the PAD and the true SPDFs are lower than 0.1 for all parameters with $\lambda < 0.9$. The significant errors only take place when the system parameters $\mathbf{\Theta}$ are near the boundary of $\mathcal{P}$, i.e., $0.9\leq \lambda \leq 1$, where the MDN has insufficient training experience to enclose $\mathbf{\Theta}$. In addition, a good proportion of system parameters with $\lambda>1$, which are excluded in the training dataset, have very low errors. It demonstrates that the MDN-based PAD has extrapolation ability to some extent.

\begin{figure}[!htb]
\center{\includegraphics[width=1\textwidth]
{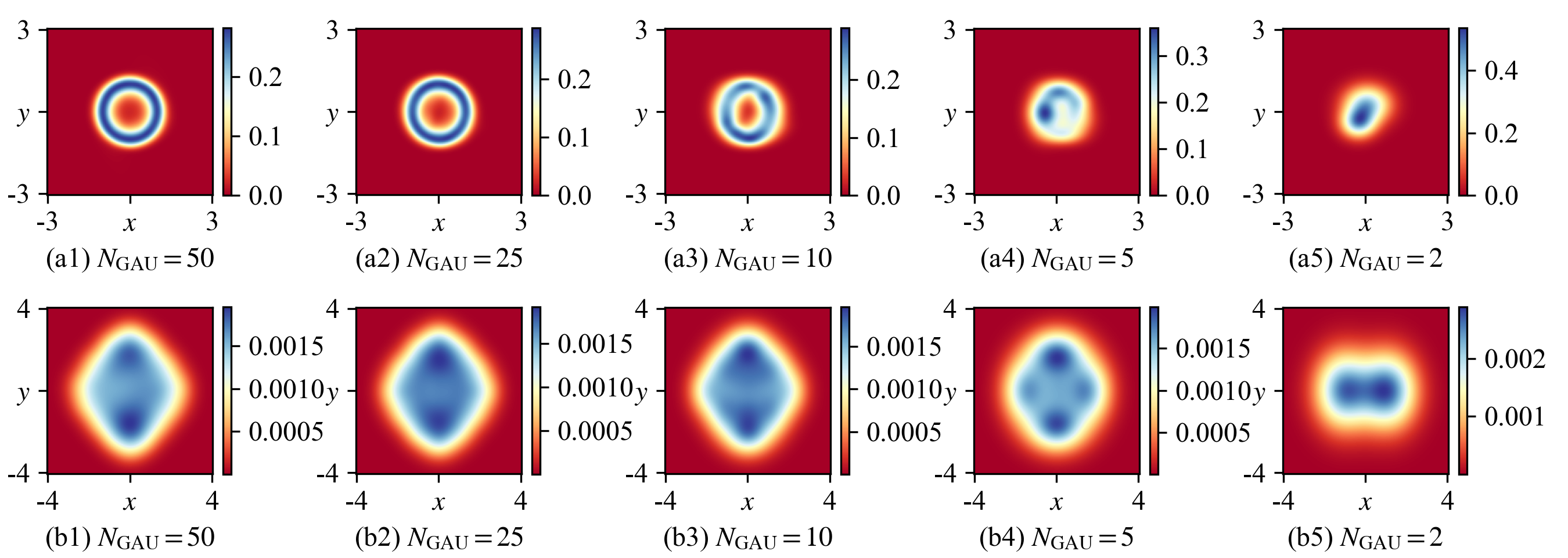}}
\caption{\label{fig:diff_num_gaussian} A smaller number of Gaussian components $N_\text{GAU}$ in the GMD reduces the learning accuracies of (a) the case (e) of the Van der Pol system in Fig.~\ref{fig:ex_van_der_pol}, Sec.~\ref{sec:ex_van_der_pol} and (b) the case (f) of the 4D system in Fig.~\ref{fig:ex_4d_sys}, Sec.~\ref{sec:ex_4d_sys}. In each subfigure, a MDN is trained for 50,000 batches and the other settings are the same as Sec.~\ref{sec:num_ex}.}
\end{figure}

A key parameter of the GMD used in the MDN-based PAD is $N_\text{GAU}$, the number of Gaussian components. Our numerical study indicates that increasing $N_\text{GAU}$ results in a better approximation, albeit at the cost of a higher requirement for GPU memory and a slower training speed. We train several MDNs with different $N_\text{GAU}$ and test them on the case (e) of the Van der Pol system in Fig.~\ref{fig:ex_van_der_pol} and the case (f) of the 4D system in Fig.~\ref{fig:ex_4d_sys}. Figure~\ref{fig:diff_num_gaussian} demonstrates that in both cases, the SPDFs can be well approximated by PADs with very limited $N_\text{GAU}$. In the first row of Fig.~\ref{fig:diff_num_gaussian}, when $N_\text{GAU}=10$, the learned SPDF of the Van der Pol system becomes significantly poor and when $N_\text{GAU}<10$, the learned SPDF cannot keep the correct ring-shaped distribution. In the second row of Fig.~\ref{fig:diff_num_gaussian}, even $N_\text{GAU}=5$ Gaussian components can learn a reasonable SPDF. We suggest that the suitable number of Gaussian components should be larger than 20, and it may vary for different systems.

\begin{figure}[!htb]
\center{\includegraphics[width=1\textwidth]
{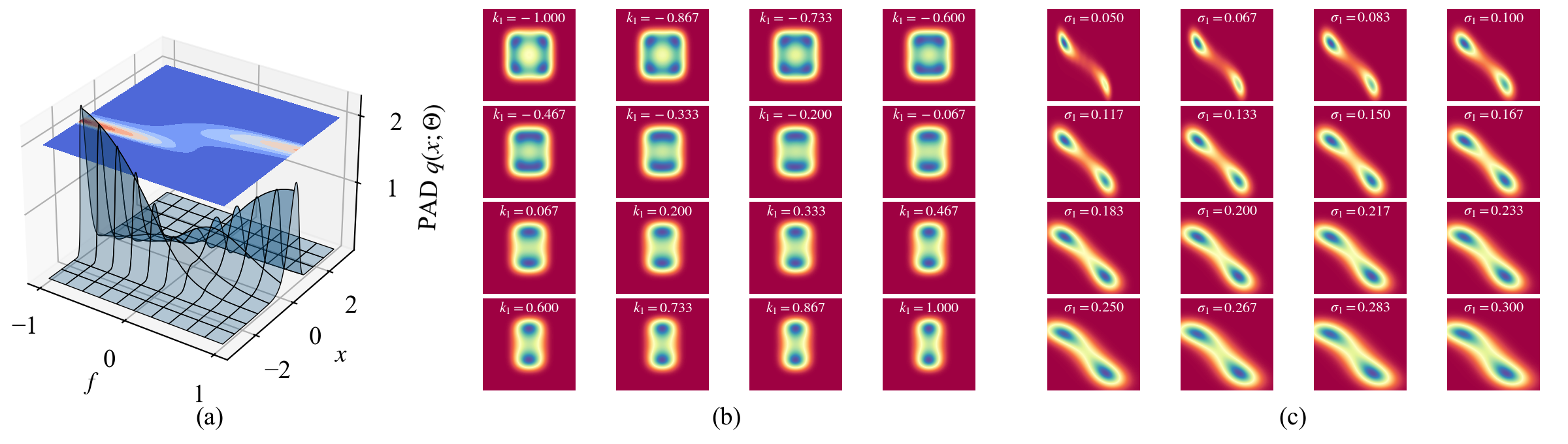}}
\caption{\label{fig:response_analysis} Efficient response analysis by using the MDN-based PAD. (a) 1000 SPDFs $q(x;\mathbf{\Theta})$ of the 1D system in Sec.~\ref{sec:sys_1d} when the parameter $f\in [-1, 1]$ varies. The other parameters are the same as Fig.~\ref{fig:tristable}(d). Each SPDF includes 1000 states in $[-3, 3]$. (b) 16 conditional SPDFs $q(x_1, x_2\in[-3, 3]|y_1=y_2=0.6;\mathbf{\Theta})$ of the 4D system in Sec.~\ref{sec:ex_4d_sys} when the parameter $k_1\in[-1,1]$ varies. The other parameters are the same as Fig.~\ref{fig:ex_4d_sys}(d). (c) 16 SPDFs $q(x,y\in[0, 1.5];\mathbf{\Theta})$ of the toggle switch system in Sec.~\ref{sec:ex_toggle_switch} when the noise intensity $\sigma_1\in[0.05, 3]$ varies. The other parameters are $a=0.25$, $b=c=1$, and $\sigma_2=0.15$. Each subfigure in (b) and (c) includes 40,000 states on a $200\times 200$ grid.}
\end{figure}

Though the training of the MDN-based PAD takes a few hours, we consider it is still fast, as a one-time training simultaneously obtains the SPDFs with all different system parameters in the parameter domain $\mathcal{P}$. When the training is finished, the PAD can be applied for  extensive high-speed response analysis even without GPU. Three examples are shown in Fig.~\ref{fig:response_analysis}. Figure~\ref{fig:response_analysis}(a) visualizes the learned SPDFs of the 1D system in Sec.~\ref{sec:sys_1d} with different system parameter $f\in[-1,1]$. When $f$ increases, the peak of the SPDF switches upwards. The simulation evaluates 1 million parameter-state pairs. It takes only 1.5 seconds on a laptop with Intel Core i7-10750H CPU. Figures~\ref{fig:response_analysis}(b) and (c) draw the evolution of the learned SPDFs of the 4D system in Sec.~\ref{sec:ex_4d_sys} and the 2D toggle switch system in Sec.~\ref{sec:ex_toggle_switch} when one of their parameters varies, respectively. As each simulation evaluates 16 SPDFs on a regular $200\times 200$ grid, both simulations evaluate $16\times 40,000=640,000$ state-parameter pairs. On the same laptop, the running times are only 1.8 and 1.3 seconds, respectively. In contrast, all the existing numerical approaches have to run a time-consuming optimization or training process to calculate the solution of every new parameter choice. Therefore, the PAD can greatly accelerate the response analysis of multi-dimensional, multi-parameter stochastic systems.

\section{Conclusion}
\label{sec:6}

In this work, we propose a MDN-based PAD to jointly solve the parameterized FPE across different choices of system parameters. This method models the stationary solution of the FPE by a convex combination of Gaussian distributions and employs a residual network to capture the intricate transformation between the system parameters and the parameters of the GMD. Through this framework, the combination weights, means, and variances of a limited number of Gaussian components adapt automatically to variations in diverse system parameters, representing the respective SPDFs within a single training process. Furthermore, we enforce the nonnegativity and normalization conditions of probability and the constraints of the GMD within the learning architecture to simplify the optimization process. Thanks to these designs, our method efficiently solves multi-dimensional, multi-parameter systems by solely considering the loss associated with the FPE, without the need for grid discretization. Numerical studies conducted on several representative systems demonstrate that our method can solve parameterized FPEs with high accuracy in a short period. The resulting PAD can be repeatedly utilized for the rapid generation of SPDFs. These findings underscore the immense potential of deep learning in accelerating the response analysis of parameter-dependent stochastic systems, especially when analytical solutions to the FPEs are challenging to obtain. However, our simulations also reveal a potential pitfall: in some cases, to prevent converging to a zero solution, it is imperative to explicitly incorporate the normalization condition into the loss function. This limitation will be the focus of our ongoing research efforts.

\appendix
\section{Numerical calculation}
\label{sec:appendix_numerical_alogrithms}

\subsection{The true SPDF}
\label{sec:app_true_SPDF}
When there is no analytical SPDF, e.g., the cases (e) and (f) in Fig.~\ref{fig:ex_4d_sys} and the cases (d), (e) and (f) in Fig.~\ref{fig:ex_6d_sys}, Monte-Carlo simulation is utilized to approximate the true conditional SPDFs $p(x_1, x_2|y_1=y_2=0.6)$ of the 4D system and $p(x_1,x_2|x_3=y_1=y_2=y_3=0)$ of the 6D system, respectively. We numerically solve these systems by the Euler-Maruyama method with the discretization step $\delta t=0.01$. The total simulation horizon is $T=200$ and we only consider the 2000 samples in $t\in[180, 200]$ to remove transients. 

For the 4D system in Sec.~\ref{sec:ex_4d_sys}, the states with both $y_1$ and $y_2$ in the interval $[0.1, 1.1]$ are gathered to approximate the condition $y_1=y_2=0.6$. For the 6D system, the states with all of $x_3$, $y_1$, $y_2$ and $y_3$ in $[-0.5, 0.5]$ are gathered to approximate the condition $x_3=y_1=y_2=y_3=0$. To obtain smooth distributions, the simulation is repeated for 10 million times with initial states uniformly sampled in the state domain $\mathcal{S}$ and the final distribution is quantified into a $400\times 400$ grid covering the $(x_1,x_2)$-plane.

As the analytical SPDF $p(\mathbf{x};\mathbf{\Theta})$ of the toggle switch system is not available either, Monte-Carlo simulation is utilized either. The setting of the Euler-Maruyama method is the same as the calculation of the previous conditional PDF. But here, the simulation is only repeated for 1 million times and the 2 billion states in $t\in[180, 200]$ are accumulated into a $400\times 400$ grid as the true SPDF.

\subsection{The L1 error between distributions}
\label{sec:app_l1_dist}
The $L_1$ error in Eq.~(\ref{eq:l1_error}) between the PAD $q$ and the true SPDF $p$ on the system parameters $\mathbf{\Theta}$ is numerically calculated by the following summation
\begin{equation}\label{eq:l1_error_approx}
d_\text{L1}(p,q;\mathbf{\Theta})\approx \sum_{\mathbf{x}\in\mathcal{G}_{\text{Error}}}|p(\mathbf{x};\mathbf{\Theta})-q(\mathbf{x};\mathbf{\Theta})|\Delta\mathbf{x},
\end{equation}
where the set $\mathcal{G}_{\text{Error}}$ includes $N_{\text{BIN}}^{D_\text{STA}}$ states on a regular grid that covers the $D_\text{STA}$-D state domain $\mathcal{S}$ and $\Delta\mathbf{x}$ is the differential element. We choose $N_\text{BIN}=1000, 200, 30$ and $10$ for the 1D, 2D, 4D and 6D systems. Therefore, the summation includes 1000, 40,000, 810,000, and 1,000,000 integral points for the four systems in Sec.~\ref{sec:sys_1d} to Sec.~\ref{sec:ex_6d_sys}, respectively.

To extensively evaluate the PAD on different system parameters, we randomly sample 10,000 groups of system parameters with analytical SPDFs. For the 1D system in Sec.~\ref{sec:sys_1d} and the 2D Van der Pol system in Sec.~\ref{sec:ex_van_der_pol}, these parameter choices are uniformly sampled in the respective parameter domains $\mathcal{P}$. The 4D system in Sec.~\ref{sec:ex_4d_sys} has the analytical SPDF only when the system parameters obey Eq.~(\ref{eq:ana_4d}). We first uniformly sample 10,000 parameter choices in $\mathcal{P}$ and then in each choice, the two noise intensities $\sigma_1$ and $\sigma_2$ are replaced by $\sqrt{ra/M}$ and $\sqrt{rb/I}$ for $r=\frac{\sigma_1^2M/a+\sigma_2^2I/b}{2}$, respectively, to satisfy Eq.~(\ref{eq:ana_4d}). Similarly, the 6D system in Sec.~\ref{sec:ex_6d_sys} has the analytical SPDF when the system parameters obey Eq.~(\ref{eq:ana_6d}). Thus the 10,000 parameter choices are uniformly sampled from the parameter domain $\mathcal{P}$ with the constraints $k_1=k_2=k_3$ and $\sigma_1=\sigma_2=\sigma_3$. The boxplots of the errors and the average errors are detailed in Fig.~\ref{fig:losses} and Tab.~\ref{tab:nn}, respectively.

\section*{Acknowledgements}

This study was partly supported by the NSF of China (Grant No. 52225211), the Key International (Regional) Joint Research Program of the NSF of China (Grant No. 12120101002), the National Natural Science Foundation of China (Grant Nos. 12202255, 12102341, and 12072264). 





\normalem
\bibliographystyle{plain}  
\bibliography{reference} 

\end{document}